\documentclass{jfm}
\usepackage{graphicx}
\usepackage{epstopdf, epsfig}

\usepackage{multirow}
\usepackage{newtxmath}
\usepackage{tikz}
\usepackage{lineno}

\definecolor{ss0}{rgb}{0.7143,0,0.2857}
\definecolor{sss0}{rgb}{0.8571,0,0.1429}
\definecolor{ssss0}{rgb}{1.0000,0,0}

\newcommand{\ddashedss}{\raisebox{2pt}{\tikz{\draw[color=ss0,dashdotted,line width = 0.9pt](0,0) -- (5mm,0);}}}
\newcommand{\ddashedsss}{\raisebox{2pt}{\tikz{\draw[color=sss0,dashdotted,line width = 0.9pt](0,0) -- (5mm,0);}}}
\newcommand{\ddashedssss}{\raisebox{2pt}{\tikz{\draw[color=ssss0,dashdotted,line width = 0.9pt](0,0) -- (5mm,0);}}}
\newcommand{\solidss}{\raisebox{2pt}{\tikz{\draw[color=ss0,solid,line width = 0.9pt](0,0) -- (5mm,0);}}}
\newcommand{\solidsss}{\raisebox{2pt}{\tikz{\draw[color=sss0,solid,line width = 0.9pt](0,0) -- (5mm,0);}}}
\newcommand{\solidssss}{\raisebox{2pt}{\tikz{\draw[color=ssss0,solid,line width = 0.9pt](0,0) -- (5mm,0);}}}

\newcommand{\dashedb}{\raisebox{2pt}{\tikz{\draw[color=blue,dashed,line width = 0.9pt](0,0) -- (5mm,0);}}}
\newcommand{\dashedv}{\raisebox{2pt}{\tikz{\draw[color=violet,dashed,line width = 0.9pt](0,0) -- (5mm,0);}}}
\newcommand{\dashedr}{\raisebox{2pt}{\tikz{\draw[color=red,dashed,line width = 0.9pt](0,0) -- (5mm,0);}}}
\newcommand{\dashedk}{\raisebox{2pt}{\tikz{\draw[color=black,dashed,line width = 0.9pt](0,0) -- (5mm,0);}}}
\newcommand{\solidk}{\raisebox{2pt}{\tikz{\draw[color=black,solid,line width = 0.9pt](0,0) -- (5mm,0);}}}
\newcommand{\solidv}{\raisebox{2pt}{\tikz{\draw[color=violet,solid,line width = 0.9pt](0,0) -- (5mm,0);}}}
\newcommand{\solidr}{\raisebox{2pt}{\tikz{\draw[color=red,solid,line width = 0.9pt](0,0) -- (5mm,0);}}}
\newcommand{\solidb}{\raisebox{2pt}{\tikz{\draw[color=blue,solid,line width = 0.9pt](0,0) -- (5mm,0);}}}

\newcommand\colcomm[1]{{#1}}

\shorttitle{Examination of outer-layer similarity over obstructing surfaces}
\shortauthor{Z. Chen and R. García-Mayoral}

\title{\colcomm{Examination of outer-layer similarity in~wall~turbulence over obstructing surfaces}}

\author{Zishen Chen\aff{1}
 \and Ricardo García-Mayoral\aff{1}
 \corresp{\email{r.gmayoral@eng.cam.ac.uk}}}

\affiliation{\aff{1}Department of Engineering, University of Cambridge, Trumpington Street, Cambridge CB2 1PZ, UK}

\begin{document}

\maketitle

\begin{abstract}
Turbulent flows over canopies of rigid filaments with different densities, $\lambda_f$, are studied using direct numerical simulations at moderate Reynolds numbers $Re_\tau\approx550-1000$. The canopies have heights $h^+\approx110-220$, and are used as an instance of obstructing substrate for the assessment of outer-layer similarity. We show that conventional methods used to determine the zero-plane displacement, $\Delta y$, can be at odds with proper outer-layer similarity and may not be applicable for flows at moderate $Re_\tau$. Instead, we determine $\Delta y$ and the length and velocity scales that recover outer-layer similarity by minimising the difference between the smooth-wall and canopy diagnostic function everywhere above the roughness sublayer, not just in the logarithmic layer. In addition, we explore the possibility of the zero-plane displacement and the friction velocity being set independently, but find that outer-layer similarity is more consistently recovered when they are coupled. We observe that although the Kármán constant, $\kappa$, may not have smooth-wall-like values, the flow statistics are smooth-wall-like in the logarithmic layer and above if the surface effect is limited within the near-wall region. This suggests a modified outer-layer similarity, where $\kappa$ is not 0.39, but turbulence is otherwise smooth-wall-like. When the canopy is dense, the flow above the tips is essentially smooth-wall-like, with smooth-wall-like $\kappa\approx0.39$ and origin essentially at the tip plane. For canopies with intermediate density, the overlying flow perceives a deeper zero-plane displacement into the canopy, which is consistent with observations reported by previous studies, but exhibits a lower Kármán constant, $\kappa\approx0.34-0.36$. For sparse canopies, $\kappa$ tends back to its smooth-wall value, and the zero-plane-displacement height is at the canopy bed. For all canopies studied, the decrease in $\kappa$ never exceeds $15\%$, which is significantly less than that obtained in some previous works using conventional methods to assess outer-layer similarity.
\end{abstract}


\section{Introduction and background}
\label{sec:intro}

Turbulent boundary layers over rough and complex surfaces are ubiquitous and are of significant environmental and industrial interest. Surface roughness can induce significant frictional drag or pressure drop for flows in engineering settings, as summarised in the reviews by \cite{flack2010review, flack2014roughness}, and \cite{chung2021predicting}. Vegetation canopies are of great ecological importance to terrestrial and aquatic ecosystems, as reviewed by \cite{finnigan2000turbulence}, \cite{belcher2012wind}, \cite{nepf2012flow, nepf2012hydrodynamics}, and \cite{brunet2020turbulent}. Porous substrates are also present in a variety of settings \citep{wood2020modeling}, such as river beds \citep{vollmer2002micro, breugem2006influence}, heat exchangers \citep{lu1998heat, dixon2012experimental}, and catalytic reactors \citep{lucci2017comparison}. In addition, engineered surfaces exposed to turbulent flows generally degrade and roughen due to erosion, fouling, and cumulative damage \citep{wu2007outer}. For these reasons, understanding the impact of complex surfaces on turbulence is essential for the modelling and control of practical flows and to improve environmental and engineering practices.

The surface topology has a direct impact on the flow within the roughness sublayer, which generally extends up to $2-3$ roughness heights, $h$, or spacings, $s$, above the roughness crests, depending on the density regime \citep{jimenez2004turbulent, macdonald2018direct, brunet2020turbulent}. Above this height, it is widely accepted that the turbulence is essentially undisturbed  and exhibits outer-layer similarity \citep{hama1954boundary, clauser1956turbulent, townsend1976structure}. The only effect is then a constant shift, $\Delta U^+$, in the mean velocity profile, while both the Kármán constant, $\kappa\approx0.39$, and the wake region remain unaffected. Experimental evidence of outer-layer similarity was provided by \cite{perry1977asymptotic} and \cite{andreopoulos1981measurements}, who reported smooth-wall-like mean-velocity profiles and turbulent statistics in the outer layer for flows over rough walls. The recovery of outer-layer similarity has also been observed in flows over a wide range of surface topologies, including 2D ribs and grooves \citep{krogstad2005experimental, leonardi2007properties, macdonald2018direct, zhang2020rough}, sand grain \citep{flack2005experimental, connelly2006velocity, amir2011turbulence, flack2023hydraulic}, prismatic roughness \citep{castro2007rough, yang2016exponential, sadique2017aerodynamic, placidi2018turbulent, sharma2020scaling, xu2021flow}, and practical rough surfaces \citep{shockling2006roughness, allen2007turbulent, wu2007outer}. \cite{jimenez2004turbulent} argued that the recovery of outer-layer similarity relies on a large scale separation, $h/\delta<1/40$, where $\delta$ is the boundary layer thickness. Numerical studies of roughness and riblets have nevertheless observed outer-layer similarity for roughness with larger blockage ratios, $h/\delta=1/8$ for cubes in an open channel and $h/\delta=1/7$ for sinusoidal roughness in a pipe \citep{leonardi2010channel, garcia2011hydrodynamic, chan2015systematic, abderrahaman2019modulation, sharma2020scaling}. As summarised in \cite{chung2021predicting}, the smooth-wall similarity in the wake region remains robust and holds even for intrusive roughness with $h/\delta\gtrsim0.15$. In this obstacle regime \citep{jimenez2004turbulent}, the protruding surface effect can completely disrupt similarity in the logarithmic layer, but the similarity is still recovered in the outer wake region \citep{flack2010review, flack2014roughness}.

Studies of wall-bounded turbulence have provided the tools for analysing and modelling rough-wall flows, with engineering models that treat roughness as a small perturbation to the smooth-wall flow \citep{flack2005experimental, flack2007examination}. However, if the roughness-induced perturbation propagates into the outer layer, the scaling based on smooth-wall similarity could \colcomm{result in inaccurate} predictions for turbulent statistics and integral quantities. Understanding the extent of roughness effects and whether smooth-wall similarity holds true is therefore of great importance to various applications. \cite{townsend1976structure} proposed the outer-layer similarity hypothesis, articulating that at a sufficiently high Reynolds number, the turbulent eddies in the outer layer would be essentially unaffected by the surface topology. The surface affects the flow only through providing the relevant scales, the wall shear stress, $\tau_w$, or the friction velocity, $u_\tau=(\tau_w/\rho)^{1/2}$, and the characteristic length scale provided by the wall-normal distance to the wall, $y$. Townsend's hypothesis is essentially a dimensional argument stating that given $\delta^+\gg1$ and $h/\delta\ll1$, surface effects are confined within the roughness sublayer, and thus the only relevant scales for the flow above are $u_\tau$ and $y$, independent of the  surface topology. Note that $u_\tau$ and $y$ are well defined for smooth-wall flows but may not be easily estimated for flows over rough and complex surfaces where the 'wall' is not obvious \citep{schultz2007rough, squire2016comparison}.

The canonical logarithmic form of the mean velocity profile is

\begin{equation}
U^+ = \frac{1}{\kappa}\log(y^++\Delta y^+)+A-\Delta U^+,
\label{eq:expect log}
\end{equation}

\noindent where $\kappa$ is the Kármán constant and $\kappa\approx0.39$ if outer-layer similarity recovers, $A$ is the log-law intercept for a smooth-wall flow, $\Delta U^+$ is the velocity deficit caused by the drag induced by surface roughness, $y^+$ is the wall-normal distance, and $\Delta y^+$ is the zero-plane displacement that recovers outer-layer similarity for the mean-velocity profile, $U^+$. The displacement $\Delta y^+$ is typically measured from the roughness tip or trough, and the zero-plane-displacement height, $y^+=-\Delta y^+$, corresponds to the height of the origin perceived by the outer-layer flow \citep{breugem2006influence, manes2011turbulent}.

Despite substantial evidence that supports Townsend's outer-layer similarity hypothesis in the presence of diverse surface topologies, some experimental studies cast doubt on its universal validity, reporting that the roughness effects can extend well into the outer layer \citep{krogstad1992comparison, krogstadt1999surface, tachie2003roughness, bhaganagar2004effect}. In these works, it was observed that the presence of roughness significantly alters the intensities of turbulent fluctuations, especially the wall-normal velocity fluctuations and Reynolds shear stress, and the mean-velocity profile even in the wake region. Additionally, recent experimental and numerical studies for turbulent flows over rough and complex surfaces, as summarised in Table \ref{tab:studies}, have reported the existence of a logarithmic layer but with values for $\kappa$, logarithmic slope, very different from the smooth-wall value, $\kappa_s\approx0.39$. Moreover, for studies with $\delta^+\approx1000-10000$ and $h/\delta\ll1$, a decrease in $\kappa$ is still observed with an increase in Reynolds number for the same roughness, suggesting an in-depth modification of the flow by the substrates. \citep{suga2010effects, manes2011turbulent, fang2018influence, okazaki2021describing, okazaki2022turbulent}. Some studies have observed that permeable roughness could lead to approximately $50\%$ drop in $\kappa$, which is more substantial than that induced by the impermeable roughness with the same geometry, implying that permeability may enhance the extent of and the intensity of roughness effects \citep{okazaki2021describing, okazaki2022turbulent, karra2022pore, esteban2022mean}. Nevertheless, the prediction of $u_\tau$, which is of great importance for the assessment of outer-layer similarity, remains a challenge in experiments \citep{chung2021predicting}. Generally, $u_\tau$ is evaluated at the zero-plane-displacement height, which is typically between the tip and trough of the obstacles, and $u_\tau$ is therefore not necessarily given by the total drag, $\tau_w$, exerted on the surface. Depending on flow conditions and apparatus, uncertainties in $u_\tau$ and turbulent statistics are typically $\sim\pm1-5\%$ \citep{schultz2007rough, schultz2013reynolds, squire2016comparison}.

\renewcommand{\arraystretch}{1.4}
\begin{table}
\centering
\begin{tabular}{lcccccc}
Studies              & Wall    & Roughness           & $\delta^+$   & $\sqrt{K^+}$ & $k_s^+$     & $\kappa$      \\[0.25cm]
B06                  & IPW     & Packed bed          & 353-678      & 0.31-9.35    & 0.11-49.6   & 0.23-0.40     \\
S10                  & IPW     & Foamed ceramics     & 150-1086$^*$ & 0.81-11.05   & 0.3-63.0    & 0.23-0.39     \\
M11                  & IPW     & Polymer foam        & 1856-5840    & 1.9-17.2     & -           & 0.31-0.33     \\
R17                  & IW      & Elastic Wall        & 180.8-307.0  & -            & -           & 0.20-0.36     \\
K17                  & APW/IPW & Pore arrays         & 120-399      & 0-11.6       & 0-19        & 0.21-0.41     \\
S20                  & APW/IPW & Spheres             & 395          & 2.62         & -           & 0.32-0.33     \\
K21                  & APW     & Spanwise rods       & 150-1137     & 3.5-25.9     & 0.4-60.9    & 0.21-0.40     \\
K22                  & APW/IW  & Spheres             & 270          & 2.56         & 2.65-6.65   & 0.32-0.35     \\
E22                  & IW      & Riblets             & 850-3840     & -            & -           & 0.30-0.37     \\
\multirow{2}{*}{F18} & IPW     & Spheres             & 862-10174    & 0.7-109      & 108-295$^*$ & 0.32-0.34$^*$ \\
                     & IW      & Hemisphere          & 986.8        & -            & 10.6        & 0.41          \\
\multirow{2}{*}{O21} & IW      & Solid ribs/grooves  & 160-2100     & -            & 642-1831    & 0.38-0.41     \\
                     & IW      & Porous ribs/grooves & 380-1730     & 1.7-16.0     & 185-1049    & 0.19-0.29     \\
\multirow{2}{*}{O22} & IW      & Solid ribs          & 220-2050     & -            & 17-1850     & 0.38-0.41     \\
                     & IW      & Porous ribs         & 240-1780     & 1.0-17.0     & 38-1340     & 0.22-0.33    
\end{tabular}
\caption{Studies that observe the modification of the logarithmic layer by the surface topology. The wall types are IPW (isotropic permeable wall), APW (anisotropic permeable wall) and IW (Impermeable wall). $\delta^+$, $\sqrt{K^+}$, $k_s^+$ and $\kappa$ are the reported friction Reynolds number ($\delta^+=\delta u_\tau/\nu$, where $\delta$ is the boundary-layer thickness or channel half-height, $u_\tau$ is the characteristic friction velocity and $\nu$ is the kinematic viscosity), permeability Reynolds number ($\sqrt{K^+}=\sqrt{K} u_\tau/\nu$ where $K$ is the permeability), roughness height ($k_s^+=k_s u_\tau/\nu$ where $k_s$ is the equivalent sand-grain roughness height) and Kármán constants, respectively. The abbreviations for the studies are B06 \citep{breugem2006influence}, S10 \citep{suga2010effects}, M11 \citep{manes2011turbulent}, R17 \citep{rosti2017numerical}, K17 \citep{kuwata2017direct}, S20 \citep{shen2020direct}, K21 \citep{kazemifar2021effect}, K22 \citep{karra2022pore}, E22 \citep{endrikat2022reorganisation}, F18 \citep{fang2018influence}, O21 \citep{okazaki2021describing}, and O22 \citep{okazaki2022turbulent}. Note that $\delta^+$ denoted by * is estimated from $k_s^+$ and $k_s/\delta$ from S10. The values for $k_s^+$ and $\kappa$ from F18, denoted by *, were provided only for some of their cases.}
\label{tab:studies}
\end{table}

Recent studies carried out by \cite{tuerke2013simulations} and \cite{lozano2019characteristic} suggest that the scaling for wall turbulence is essentially local, and is set by the local mean shear and production rate of turbulent kinetic energy, with no explicit reference to the wall-normal distance, $y$. This implies that the traditional scaling based on $y$ and $u_\tau$ happens to hold because of the one-to-one correspondence between the latter and the local production and shear, but this correspondence does not need to hold necessarily for flows over non-smooth walls. As part of this work, we investigate, for flows that exhibit an apparent loss of outer-layer similarity, whether the local scale can still have correspondence to a friction velocity, $u_\tau^\star$, and a length scale, $y_*$, where $y_*$ is the wall-normal distance to the zero-plane-displacement height, $y_*=0$, but $u_\tau^\star$ is not necessarily evaluated at $y_*=0$. In this work, superscript $(\cdot)^\star$ denotes wall units defined by $\nu$ and $u_{\tau}^\star$ decoupled from $y_*=0$, and superscript $(\cdot)^+$ denotes wall units defined by $\nu$ and $u_{\tau}^*$ evaluated at $y_*=0$. Subscript $(\cdot)_*$ denotes outer units that are normalised by the bulk velocity, $U_b$, and outer length scale, $y_*$. 

The diagnostic function of the mean-velocity profile in equation (\ref{eq:expect log}) is 

\begin{equation}
\beta=y_*^+\frac{\partial U^+}{\partial y_*^+},
\label{eq:expect diag}
\end{equation}

\noindent where $y_*=(y+\Delta y)/(\delta+\Delta y)$ is the wall-normal distance from the zero-plane-displacement height, at $y_*=0$, which would exhibit a plateau $\beta\approx1/\kappa$ in the logarithmic layer, if outer-layer similarity recovers \citep{mizuno2011mean, luchini2018structure}. This diagnostic function is useful because deviations from the log-law profile are typically more apparent in $\beta$ in equation (\ref{eq:expect diag}) than in $U^+$ in equation (\ref{eq:expect log}). Many previous studies therefore rely on the existence of this plateau in $\beta$ to determine the extent of the logarithmic layer and the inner scaling for flows over roughness \citep{breugem2006influence, suga2010effects}. Particularly, the linear relation between $U^+$ and $\log(y_*^+)$ in equation (\ref{eq:expect log}) is enforced by choosing a $\Delta y$ that yields a plateau in $\beta(y_*^+)$. The inner velocity and length scales are then determined based on $u_\tau^*$ evaluated at the reference height, $y_*=0$, yielding values for $\kappa$ that are not necessarily smooth-wall like, as listed in Table \ref{tab:studies}. However, a logarithmic layer with a plateau in $\beta$ emerges only in flows at very high $Re_\tau$ \citep{lee2015direct, hoyas2022wall}.  More importantly, outer-layer similarity, by definition, refers to the similarity in not just the logarithmic layer but also the wake, the whole outer region. In the present work, we argue that for flows at all but the highest $Re_\tau$, neglecting smooth-wall similarity in the wake region while enforcing a plateau in the diagnostic function could result in \colcomm{spurious} predictions of parameters, including $\Delta y$, $u_\tau^*$ and $\kappa$, and friction-scaled turbulent statistics. In this study, we determine the zero-plane displacement, $\Delta y$, by minimising the deviation compared to a smooth-wall flow of the diagnostic function not only in the logarithmic layer but also above. We assess the validity as a scaling velocity of the friction velocity, $u_\tau^*$ and $u_\tau^\star$, both measured at the height of zero-plane displacement and set as an independent, free parameter. Additionally, we examine whether the value of $\kappa$ is modified by the type of surface or not. We probe the existence of outer-layer similarity in an extensive dataset of canopy flows. This choice is motivated by canopies being an instance of porous-like complex surfaces which are particularly obstructing and intrusive to the flow \citep{ghisalberti2009obstructed}.

The paper is organised as follows. The numerical method and relevant canopy parameters are presented in \S \ref{sec:dns}. Results, with particular emphasis on scaling for the outer-layer turbulence, are discussed in \S \ref{sec:Results and discussion}. Finally, the conclusions are summarised in \S \ref{sec:Conclusions}.

\section{Direct numerical simulations}
\label{sec:dns}

\begin{figure}
\vspace*{-2mm}
\centering
\includegraphics[width=0.8\textwidth]{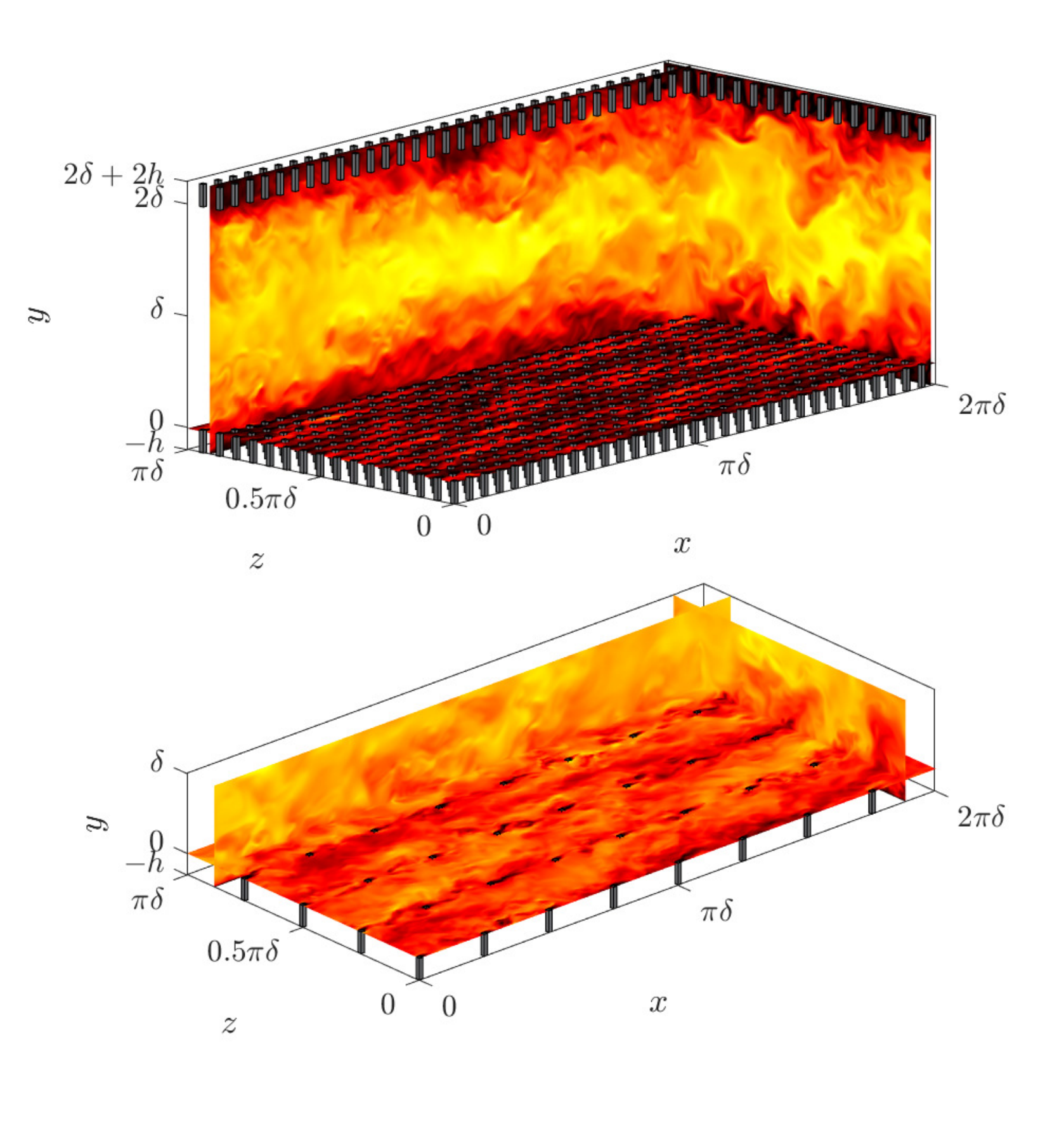}
\put (-110mm,105mm) {$(a)$}
\put (-110mm,50mm) {$(b)$}
\vspace*{-8mm}
\caption{Schematics of the numerical domain of $(a)$ full-channel case C108$_{550}$ and $(b)$ open-channel case O400$_{550}$. An instantaneous realisation of the streamwise velocity is shown in the orthogonal planes.}
\label{fig:uncontour}
\end{figure}

We present results for a series of direct-numerical simulations (DNS) of closed and open channels with canopies of rigid filaments covering the walls at moderate Reynolds numbers, $Re_\tau\approx500-1000$. We note that these $Re_\tau$ are sufficiently high for convective effects to be dominant, such that the interpretation of the turbulent statistics in these canopy flows may be extrapolated to cases with higher $Re_\tau$ \citep{sharma2020scaling}. The streamwise, spanwise, and wall-normal directions are $x$, $z$, and $y$, respectively. A schematic of the numerical domain is portrayed in figure \ref{fig:uncontour}. The dimensions of the closed channels are $L_x \times L_z \times L_y=2\pi\delta \times \pi\delta \times 2(\delta+h)$, where $h$ is the canopy height and $\delta=1$ is the distance between the channel centre and the canopy-tip planes. The canopy region is below $y=0$ for the bottom wall, and above $y=2\delta$ for the top wall. This domain size is large enough to reproduce the one-point statistics for the friction Reynolds numbers considered in this study, without imposing artificial constraints on the largest turbulent eddies \citep{lozano2014effect}.

We vary the canopy density by changing the spacing between elements, resulting in frontal densities $\lambda_f\approx0.01-2.04$, defined as the ratio between the frontal area of the obstacles and the total plan area. This covers a broad range from sparse to dense canopies based on the notional limit $\lambda_f\approx0.1$ proposed by \cite{nepf2012flow}. All canopies in the closed channel consist of collocated prismatic posts with thickness $\ell_x^+=\ell_z^+\approx24$ and height $h^+\approx110$. Relevant simulation parameters are listed in table \ref{tab:parm}. For the canopy simulations, letters C and O denote closed and open channels, and the number that follows denotes the approximate spacing, $s^+=L_x^+/n_x=L_z^+/n_z$, between the canopy elements, where $n_x$ and $n_z$ are the numbers of elements in the streamwise and spanwise directions, respectively. The number in the subscript is the approximate friction Reynolds number of the flow. Cases C216$_{900}$, C288$_{900}$, and C432$_{900}$ conducted at $Re_\tau\approx900$ match the geometry parameters, $s^+$, $l^+$ and $h^+$, of the sparse and intrusive cases C216$_{550}$, C288$_{550}$, and C432$_{550}$ in inner units, respectively. These cases at high Reynolds numbers are conducted to verify that outer-layer similarity can recover, even for flows over intrusive textures, provided a large enough core flow unperturbed by the roughness. Cases C$_{550}$, C$_{600}$, C$_{900}$, O$_{550}$, and O$_{1000}$ are reference smooth-wall simulations.

DNSs of sparse canopies, $\lambda_f\approx0.01$, from \cite{sharma2020scaling} are included for the assessment of outer-layer similarity in open-channel flows. These channels are bounded by a bottom no-slip wall and a top free-slip surface at $y=\delta$, as shown in figure \ref{fig:uncontour}$(b)$. Case O400$_{550}$ consists of prismatic posts with sides $\ell_x^+=\ell_z^+\approx20$, and height $h^+\approx110$. The canopy of O400$_{1000}$ matches the dimensions of O400$_{550}$ in inner units, with thickness $\ell_x^+=\ell_z^+\approx20$, and height $h^+\approx110$, while the canopy of C800$_\text{1000}$ matches the dimensions of O400$_{550}$ in outer units, with thickness $\ell_x/\delta=\ell_z/\delta\approx0.04$, and height $h/\delta\approx0.2$. The reference friction velocity, $u_\tau$, in table \ref{tab:parm} is calculated from the total shear stress at the canopy tips for the full-channel cases and from the net drag for the open-channel cases. This is the reference friction velocity used in $Re_\tau=\delta u_\tau/\nu$ and in the other friction-scaled variables discussed in this section.

\renewcommand{\arraystretch}{1.4}
\setlength{\tabcolsep}{6pt}
\begin{table}
\centering
\begin{tabular}{lccccccccc}
 &
  Case &
  $Re_\tau$ &
  $\lambda_f$ &
  $N_x\times N_z$ &
  $s/h$ &
  $\Delta x^+$ &
  $\Delta z^+$ &
  $\Delta y_f^+$ &
  $\Delta y_t^+$ \\[0.25cm]
\multirow{14}{*}{Channel flow} &
  C$_{550}$ &
  550.6 &
  - &
  - &
  - &
  9.01 &
  4.50 &
  0.27 &
  - \\
 & C$_{600}$     & 603.3  & -    & -             & -    & 8.77 & 4.39 & 0.29 & -    \\
 & C$_{950}$     & 950.2  & -    & -             & -    & 7.77 & 3.89 & 0.32 & -    \\
 & C36$_{550}$   & 550.3  & 2.04 & $96\times 48$ & 0.33 & 2.00 & 2.00 & 2.95 & 0.50 \\
 & C54$_{550}$   & 550.1  & 0.91 & $64\times 32$ & 0.49 & 3.00 & 3.00 & 2.53 & 0.50 \\
 & C72$_{550}$   & 551.9  & 0.51 & $48\times 24$ & 0.65 & 4.01 & 4.01 & 2.01 & 1.00 \\
 & C108$_{550}$  & 548.5  & 0.23 & $32\times 16$ & 0.98 & 5.98 & 2.99 & 2.00 & 1.00 \\
 & C144$_{550}$  & 547.0  & 0.13 & $24\times 12$ & 1.31 & 5.97 & 2.98 & 1.19 & 1.19 \\
 & C216$_{550}$  & 549.0  & 0.06 & $16\times 8$  & 1.96 & 5.99 & 2.99 & 1.00 & 1.00 \\
 & C288$_{550}$  & 547.9  & 0.03 & $12\times 6$  & 2.62 & 5.98 & 2.99 & 0.79 & 0.79 \\
 & C432$_{550}$  & 548.9  & 0.01 & $8\times 4$   & 3.93 & 5.99 & 2.99 & 0.60 & 0.60 \\
 & C216$_{900}$  & 895.1  & 0.06 & $32\times 16$ & 1.96 & 8.26 & 4.13 & 1.1  & 1.1  \\
 & C288$_{900}$  & 895.7  & 0.03 & $24\times 12$ & 2.62 & 8.26 & 4.13 & 1.1  & 1.1  \\
 & C432$_{900}$  & 892.4  & 0.01 & $16\times 8$  & 3.93 & 8.26 & 4.13 & 1.1  & 1.1  \\[0.25cm]
\multirow{5}{*}{\begin{tabular}[c]{@{}l@{}}Open-channel \\ flow\end{tabular}} &
  O$_{550}$ &
  537.1 &
  - &
  - &
  - &
  8.79 &
  4.39 &
  0.21 &
  - \\
 & O$_{1000}$    & 994.6  & -    & -             & -    & 8.14 & 4.07 & 0.32 & -    \\
 & O400$_{550}$  & 529.7  & 0.01 & $8\times 4$   & 3.65 & 4.33 & 4.33 & 0.20 & 1.77 \\
 & O400$_{1000}$ & 1062.4 & 0.01 & $16\times 8$  & 3.53 & 4.35 & 4.35 & 0.35 & 2.76 \\
 & O800$_{1000}$ & 1001.8 & 0.01 & $8\times 4$   & 3.61 & 4.10 & 4.10 & 0.33 & 3.45
\end{tabular}
\caption{Simulation parameters: $Re_\tau=\delta u_\tau/\nu$ is the friction Reynolds number based on $\nu$, $\delta$ and $u_\tau$ evaluated at the canopy tips; $\lambda_f$ is the frontal density; $N_x$ and $N_z$ are the numbers of canopy elements in the streamwise and spanwise directions, respectively; $h$ and $s$ are the canopy height and the spacing in the streamwise and spanwise directions; $\Delta x^+$ and $\Delta z^+$ are the streamwise and spanwise resolutions; $\Delta y_f^+$ and $\Delta y_t^+$ are the wall-normal resolutions at the floor and canopy tips. The wall-normal resolution is validated in Appendix for reference. O$_{550}$ and open-channel canopy cases are from \cite{sharma2020scaling}.}
\label{tab:parm}
\end{table}

The DNS code implemented in this study is from \cite{sharma2020scaling, sharma2020turbulent_b} and has been validated in \cite{sharma2020turbulent}. It is summarised here for reference. The numerical method solves the three-dimensional incompressible Navier-Stokes equations,

\begin{align}
  \frac{\partial\mathbf{u}}{\partial{t}} + \mathbf{u}\cdot\nabla\mathbf{u} &= -\nabla{p} + \frac{1}{Re}\nabla^2\mathbf{u},\label{eq:ns}  \\ 
  \nabla\cdot\mathbf{u} &= 0,
\end{align}

\noindent where $\mathbf{u}$ is the velocity vector $\langle u,w,v \rangle$ with components in the streamwise, spanwise and wall-normal directions, respectively, $p$ is the kinematic pressure, and $Re$ denotes the bulk Reynolds number $Re=U_b\delta/\nu$ based on $U_b$, $\delta$, and the kinematic viscosity, $\nu$. No-slip and no-penetration boundary conditions are enforced at both walls. The canopy elements are explicitly resolved using a direct-forcing, immersed-boundary method \citep{iaccarino2003immersed, garcia2011hydrodynamic}. The numerical domain is periodic in the wall-parallel directions, which are discretised spectrally. A second-order central difference scheme on a staggered grid is used in the wall-normal direction to avoid the 'chequerboard' problem \citep{ferziger2002computational}. The wall-normal grid is stretched with $\Delta y_{max}^+\approx4.5$ at the channel centre for the closed-channel simulations. For the open-channel simulations, $\Delta y_{max}^+\approx2.2$ when $Re_\tau\approx550$ and $\Delta y_{max}^+\approx5.3$ when $Re_\tau\approx1000$. $\Delta y_{min}^+$ occurs at the floor or tips, wherever the mean shear is the highest; $\Delta y_{min}^+\approx0.5-1$ is at the tips for the intermediate to dense canopies ($\lambda_f\gtrsim$0.1), and $\Delta y_{min}^+\approx0.3-0.8$ is at the floor for the sparse canopies ($\lambda_f\lesssim$0.1). The wall-normal grid resolutions are listed in table \ref{tab:parm}.

The typical wall-parallel resolutions are $\Delta x^+\lesssim8$ and $\Delta z^+\lesssim4$ for the DNS of smooth-wall turbulent flows \citep{jimenez1991minimal}. However, for the filament canopies considered in this study, the element-induced eddies are typically of the order of or smaller than the element thickness \citep{poggi2004effect}. Therefore, the wall-parallel grids are smaller than $\ell_x^+$ and $\ell_z^+$ to resolve the eddies induced by the canopy elements, as presented in table \ref{tab:parm}. To resolve the turbulence within and above the roughness sublayer without inducing excess computational cost, the numerical domain is partitioned into blocks with different wall-parallel resolutions \citep{garcia2011hydrodynamic}. The blocks that contain the roughness sublayer have a more refined resolution than the block encompassing the channel centre. In the fine blocks, the grid resolution resolves not just the turbulent scales but also the canopy geometry and element-induced eddies. The height of these blocks is chosen such that the small and rapid element-induced eddies naturally diffuse and damp out before reaching the coarse block at the channel centre, which has a standard $\Delta x^+\approx8$ and $\Delta z^+\approx4$ resolution. This is verified \emph{a posteriori} by examining the spectral densities of turbulent fluctuations near the interface to ensure that any small-wavelength signal has already vanished.

The time advancement uses a fractional-step method with a three-substep Runge-Kutta scheme where pressure is corrected to enforce incompressibility \citep{le1991improvement, perot1993analysis},

\begin{align}
\left[\text{I}-\Delta t\frac{\beta_k}{Re}\text{L}\right]\mathbf{u}^n_k &= \mathbf{u}^n_{k-1}+\Delta t\left[\frac{\alpha_k}{Re}\text{L}\mathbf{u}^n_{k-1}-\gamma_k\text{N}\mathbf{u}^n_{k-1}-\zeta_k\text{N}\mathbf{u}^n_{k-2}-\left(\alpha_k+\beta_k\right)\text{G}p^n_k\right],\\ \label{eq:ns_dis1}
\text{DG}\phi^n_k &= \frac{1}{\Delta t\left(\alpha_k+\beta_k\right)}\text{D}\mathbf{u}^n_k,\\
\mathbf{u}^n_{k+1}&=\mathbf{u}^n_k-\Delta t\left(\alpha_k+\beta_k\right)\text{G}\phi^n_k,\\
p^n_{k+1}&=p^n_k+\phi^n_k,
\end{align}

\noindent where $k = 1,2,3$ are the Runge-Kutta substeps (e.g. $u^0_0=u^0$, $u^0_3=u^1$), $\Delta t$ is the time step, $\text{I}$ is the identity matrix, $\text{L}$, $\text{G}$ and $\text{D}$ are the discretised Laplacian, gradient and divergence operators, $\text{N}$ is the advective term dealiased with the 2/3-rule \citep{canuto2012spectral}, and $\alpha_k$, $\beta_k$, $\gamma_k$ and $\zeta_k$ are the integration coefficients adapted from \cite{le1991improvement}. The channel is driven by a constant mean pressure gradient, with the flow rate adjusted to obtain the targeted friction Reynolds number. Each simulation is run for at least 10 largest-eddy-turnover times, $\delta/u_\tau$, to wash out any initial transients. Once the flow reaches a statistically steady state, statistics are collected over another $20\delta/u_\tau$.

\section{Results and discussion}
\label{sec:Results and discussion}

In this section, we present and discuss the scaling for the outer-layer flow, aiming to show that for canopy flows that exhibit an apparent loss of outer-layer similarity, a modified outer-layer similarity can be recovered when using the appropriate velocity and length scales.

\subsection{Depth of roughness layer}
\label{subsec:Extent}

Before we set out to investigate outer-layer similarity, it is important to establish a lower bound for $y$ from which it can be expected to hold. In the immediate vicinity of a complex surface, the flow cannot be expected to be universal, but specific to the particular surface topology. Outer-layer similarity should not be expected within the roughness sublayer, where turbulence is directly perturbed by the element-induced flow. Thus, we first need to identify the height above which the direct effect of the texture, manifesting as a texture-coherent signature in the flow field, vanishes effectively. The height of the roughness sublayer assumed here to be the height beyond which the element-induced flow vanishes, is generally a function of the element spacing or height, depending on the density regime \citep{jimenez2004turbulent, brunet2020turbulent}. On the basis of canopy geometry and configuration, the frontal density $\lambda_f$ gives a notional measure of canopy density \citep{wooding1973drag, nepf2012flow}. For conventional sparse canopies ($\lambda_f\lesssim0.1$) with element spacing larger than height, the roughness sublayer thickness is typically a function of the canopy height \citep{poggi2004effect, flack2007examination, abderrahaman2019modulation, sharma2020scaling}. However, for dense canopies ($\lambda_f\gtrsim0.5$), the flow within the obstacles is 'sheltered' from the turbulent flow as the elements interact with turbulence only in the vicinity of the tips, and thus the height of the roughness sublayer depends on the element spacing \citep{macdonald2018direct, placidi2018turbulent, sharma2020turbulent_b}. In the very dense limit, where the element spacings are vanishingly small, the eddies are essentially precluded from penetrating within the texture, and the overlying flow essentially perceives a smooth wall at the tips \citep{sharma2020turbulent_b, brunet2020turbulent}. 

To measure the extent of roughness effects, here we quantify the intensity of the element-induced flow using standard triple decomposition \citep{reynolds1972mechanics}

\begin{align}
    \mathbf{u}(x,y,z,t) &= \mathbf{U}(y)+\mathbf{u'}(x,y,z,t),\\
    \mathbf{u'}(x,y,z,t)&= \tilde{\mathbf{u}}(x,y,z)+\mathbf{u''}(x,y,z,t),
\end{align}

\noindent where $\mathbf{u}$ is the full instantaneous velocity vector field $\langle u,w,v \rangle$, $\mathbf{U}$ is the mean velocity profile, and $\mathbf{u'}$ is the full temporal and spatial turbulent fluctuation, decomposed into a time-averaged but spatially varying component, $\tilde{\mathbf{u}}$, and the remaining time-varying fluctuation, $\mathbf{u''}$. $\mathbf{U}$ is the velocity averaged in time and in the wall-parallel directions, and $\tilde{\mathbf{u}}$, often termed the dispersive flow \citep{castro2021channel, modesti2021dispersive}, is obtained from the average of the flow in time only. Therefore, the r.m.s. of $\tilde{\mathbf{u}}$ at each height gives a measure of the intensity of the coherent spatial fluctuation induced by the canopy elements. \cite{abderrahaman2019modulation} argued that $\tilde{\mathbf{u}}$ does not contain the whole element-coherent signal, but it nevertheless gives a good measure of its intensity.

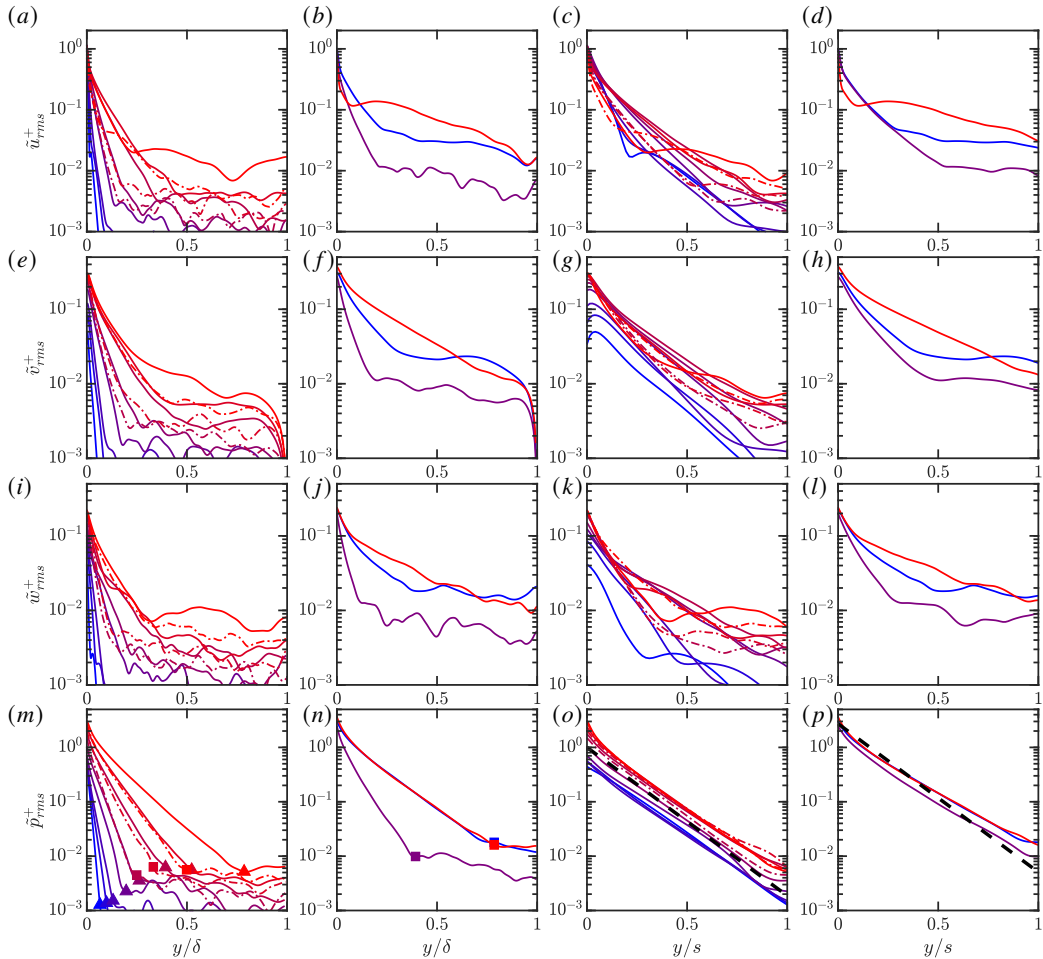
\begin{figure}
\centering
\includegraphics[width=\textwidth]{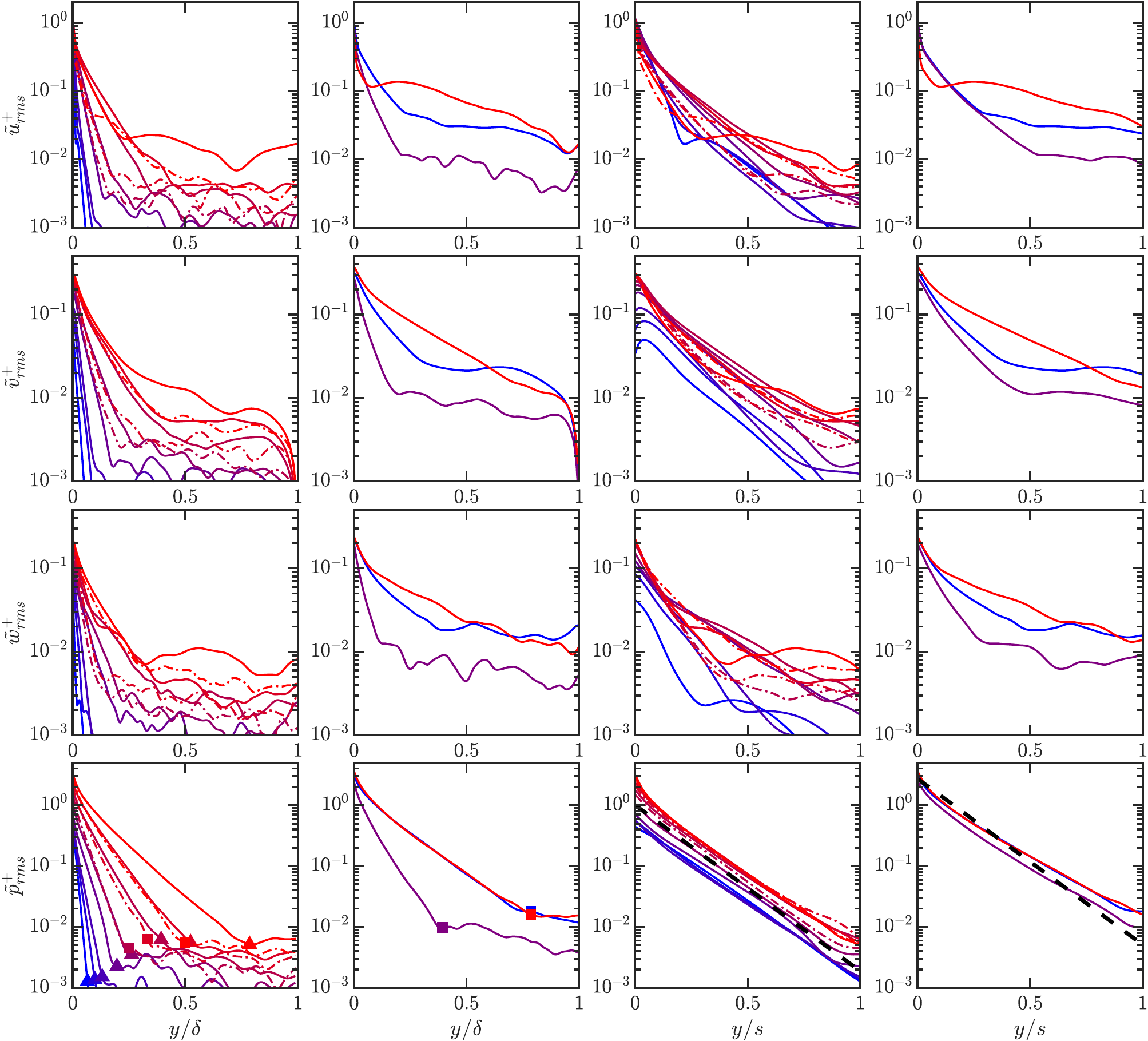}
\put (-137mm,123mm){$(a)$}
\put (-98mm,123mm) {$(b)$}
\put (-65mm,123mm) {$(c)$}
\put (-32mm,123mm) {$(d)$}
\put (-137mm,91mm) {$(e)$}
\put (-98mm,91mm)  {$(f)$}
\put (-65mm,91mm)  {$(g)$}
\put (-32mm,91mm)  {$(h)$}
\put (-137mm,61mm) {$(i)$}
\put (-98mm,61mm)  {$(j)$}
\put (-65mm,61mm)  {$(k)$}
\put (-32mm,61mm)  {$(l)$}
\put (-137mm,31mm) {$(m)$}
\put (-98mm,31mm)  {$(n)$}
\put (-65mm,31mm)  {$(o)$}
\put (-32mm,31mm)  {$(p)$}
\caption{R.m.s. velocity and pressure fluctuations of the element-induced, dispersive flow normalised by $u_\tau$ evaluated at the canopy tips. $(a,e,i,m)$ and $(c,g,k,o)$, full-channel cases; $(b,f,j,n)$ and $(d,h,l,p)$, open-channel cases. The solid lines from blue to red are cases C36$_{550}$ to C432$_{550}$ in the first and third columns and cases O400$_{550}$, O400$_{1000}$, and O800$_{1000}$ in the second and fourth columns; the dashed-dotted lines \protect\ddashedss, \protect\ddashedsss and \protect\ddashedssss are cases C216$_{900}$, C288$_{900}$ and C432$_{900}$ that have similar geometry parameters in inner units to those of cases C216$_{550}$ \protect\solidss, C288$_{550}$ \protect\solidsss, and C432$_{550}$ \protect\solidssss. \colcomm{The square markers represent height $y=s$ for cases at $Re_\tau\approx550$, and the triangle markers represent that for those at $Re_\tau\approx900-1000$.}}
\label{fig:cond_stats}
\end{figure}

As shown in figure \ref{fig:cond_stats}, the intensity of the element-induced fluctuations decays exponentially with $y$ above the tips. A similar decaying pattern has been observed in flows over superhydrophobic surfaces \citep{seo2015pressure}, three-dimensional sinusoidal roughness \citep{chan2018secondary}, prismatic roughnesses \citep{abderrahaman2019modulation}, and filament canopies \citep{sharma2020turbulent_b}. The cause can be traced to the pressure in this region satisfying a Laplace equation with two components, one forced by the nonlinear terms of the overlying flow, which is essentially texture-incoherent, and one forced by the effective boundary conditions at $y=0$, induced by the texture. The latter then takes the form $\sim e^{-y/\lambda}$ for each excited wavelength, for which the first texture harmonic, $\lambda=s$, decays more slowly and dominates \citep{kamrin2010effective, seo2015pressure}. The velocities satisfy in turn their own corresponding Laplace equations, with additional source terms from this texture-induced pressure, leading to similar exponential decays. Figure \ref{fig:cond_stats} evidences this exponential decay with $y/s$, which is particularly clear for the pressure, as well as for the wall-normal velocity, and to a lesser extent for the tangential velocities, which is to be expected given their more intense source terms in their respective Laplace equations.

For all canopies considered, the texture-coherent pressure and velocity fluctuations essentially vanish at approximately one canopy spacing above the tips, as depicted in figure \ref{fig:cond_stats}. However, this implies that the sparse ($\lambda_f\lesssim0.1$) and tall ($h\approx0.2\delta$) canopies, C216$_{550}$, C288$_{550}$, C432$_{550}$, O400$_{550}$ and O800$_{1000}$, are significantly more intrusive to the overlying flow compared to the other canopies with either intermediate to high density ($\lambda_f\gtrsim0.1$) or small height ($h\approx0.1\delta$), as their element-induced flows penetrate into the channel as far as $y\approx0.5\delta$, or even beyond. This implies that the roughness sublayer of these intrusive canopies can extend well into the overlying flow and reach the channel centre. \cite{sharma2020turbulent_b} have reported a similar behaviour for the element-induced flow over dense canopies, for which the element-induced velocity fluctuations become negligible at one canopy spacing above the tips regardless of the canopy height. However, their element-induced flows caused a more profound modification of the background turbulence, which only became smooth-wall-like at heights $y/s>2-3$ above the tips. For the present flows, however, it will be demonstrated in \S \ref{subsec:Sens} and \S \ref{subsec:Univ} that, for the cases which exhibit it, outer-layer similarity recovers above a roughness sublayer that extends only to a height $y/s\approx 1$ above the tips.

\subsection{Logarithmic velocity profiles over smooth and rough walls}\label{subsec:Log}
In this section, we discuss and appraise the conventional methods used to assess the existence of a logarithmic layer and find the zero-plane-displacement height, $y_*=0$, that sets the velocity and length scales, $u_\tau^*$ and $y_*$, for turbulent flows over roughness. Their values are generally determined by using the total drag \citep{jackson1981displacement, raupach1992drag, cheng2007flow, leonardi2010channel, squire2016comparison}, by fitting $U^+$ to be proportional to $\log(y_*^+)$ in the logarithmic layer \citep{clauser1956turbulent, flack2014roughness}, or by, equivalently, enforcing a plateau in $\beta$ \citep{breugem2006influence, suga2010effects, manes2011turbulent}.

In experiments, friction velocity and zero-plane-displacement height are generally estimated based on the total drag exerted on the surface, as summarised in \cite{chung2021predicting}. For small and sparsely distributed roughness, where the overlying flow penetrates all the way to the floor, this method generally yields $u_\tau^*$ and $\Delta y$ that recover outer-layer similarity \citep{flack2005experimental, wu2007outer, schultz2013reynolds}. However, for dense and tall roughness, the total drag method could result in nonphysical prediction for both $u_\tau^*$ and $\Delta y$. The element-induced drag is significant for dense roughness, and therefore the point of action of the total drag, $\Delta y=\int_h D(y)ydy/\int_h D(y)dy$ \citep{jackson1981displacement}, is located at an intermediate point in the roughness sublayer. However, DNSs of dense filament canopies have illustrated that the zero-plane-displacement height approaches the tips as the overlying turbulence interacts only with the upper part of the obstacles and cannot perceive the floor \citep{sharma2020turbulent_b, brunet2020turbulent}. In turn, for sparse canopies, the zero-plane-displacement height approaches the floor, even when most of the drag is still exerted by the canopy elements \citep{sharma2020scaling}. Consequently, the total drag may not \colcomm{necessarily} be directly relevant for the assessment of outer-layer similarity and the estimates of the zero-plane displacement, $\Delta y$, and the friction velocity evaluated at $y_*=0$, $u_\tau^*$.

\begin{figure}
\centering
\vspace*{3mm}
\includegraphics[width=\textwidth]{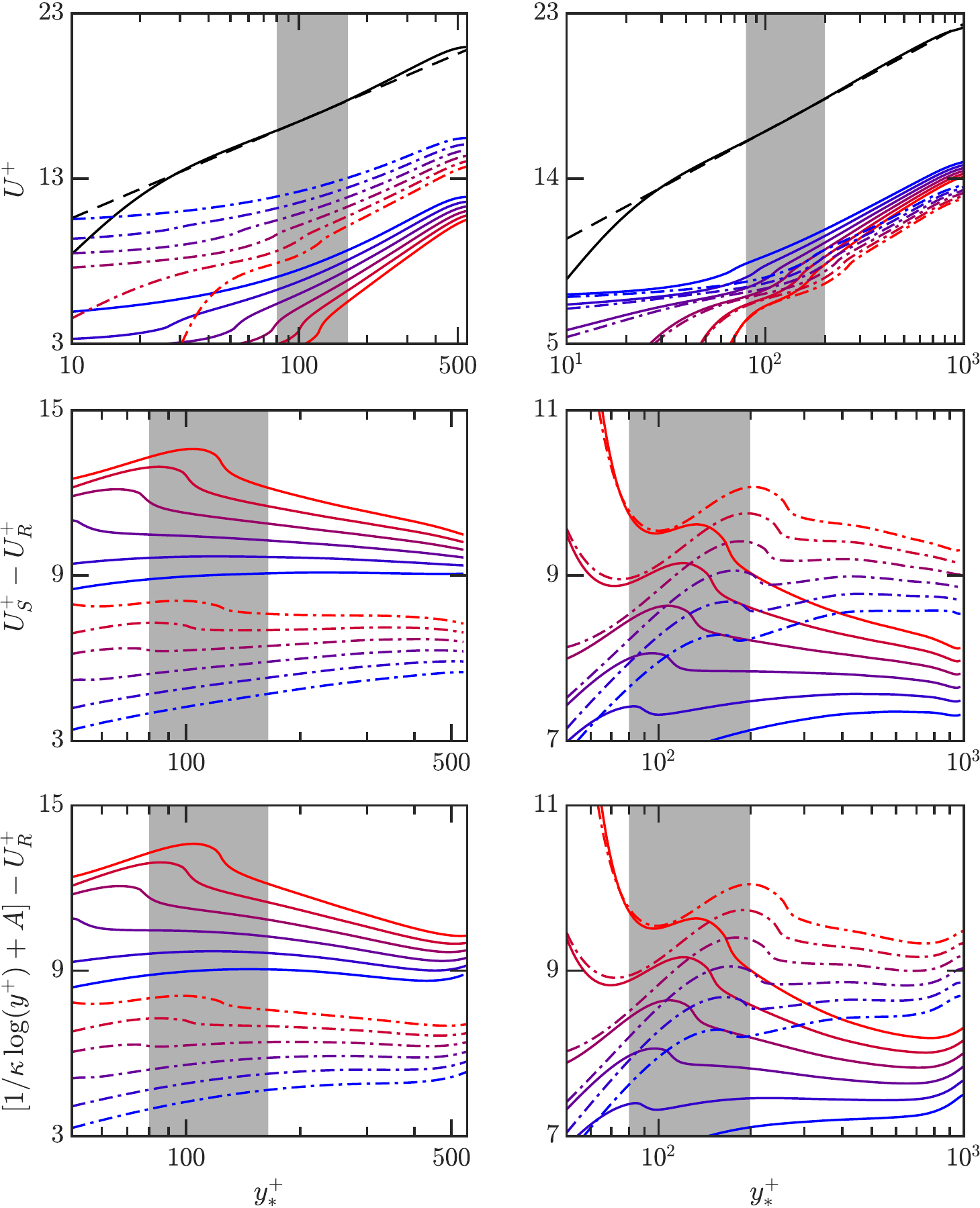}
\put (-136mm,164mm){$(a)$}
\put (-68mm,164mm) {$(b)$}
\put (-136mm,109mm){$(c)$}
\put (-68mm,109mm) {$(d)$}
\put (-136mm,55mm) {$(e)$}
\put (-68mm,55mm)  {$(f)$}
\caption{Mean velocity and velocity deficit profiles for $(a,c,e)$ cases C144$_{550}$ (solid colour lines) and C432$_{550}$ (dashed colour lines), and $(b,d,f)$ cases O400$_{1000}$ (solid colour lines) and O800$_{1000}$ (dashed colour lines). From blue to red, results are based on $(a, c)$ $\Delta y=0$ to $0.25\delta$ and $(b, d)$ $\Delta y=0.05\delta$ to $0.15\delta$; the black solid lines are reference smooth-wall profiles $(a)$ C$_{550}$ and $(b)$ O$_{1000}$; the black dashed lines are the smooth-wall log-law profile, $1/\kappa\log(y^+)+A$; and the shaded area marks the logarithmic region for smooth-wall flows, $y^+=80$ to $y=0.3\delta$ for closed channels and $y^+=80$ to $y=0.2\delta$ for open channels. The upper bound of the log region in open channels is lower than that in a closed channel because the free-slip surface induces a 'cutoff' to the wake region, which limits the extent of the log layer.}
\label{fig:rough_delU}
\end{figure}
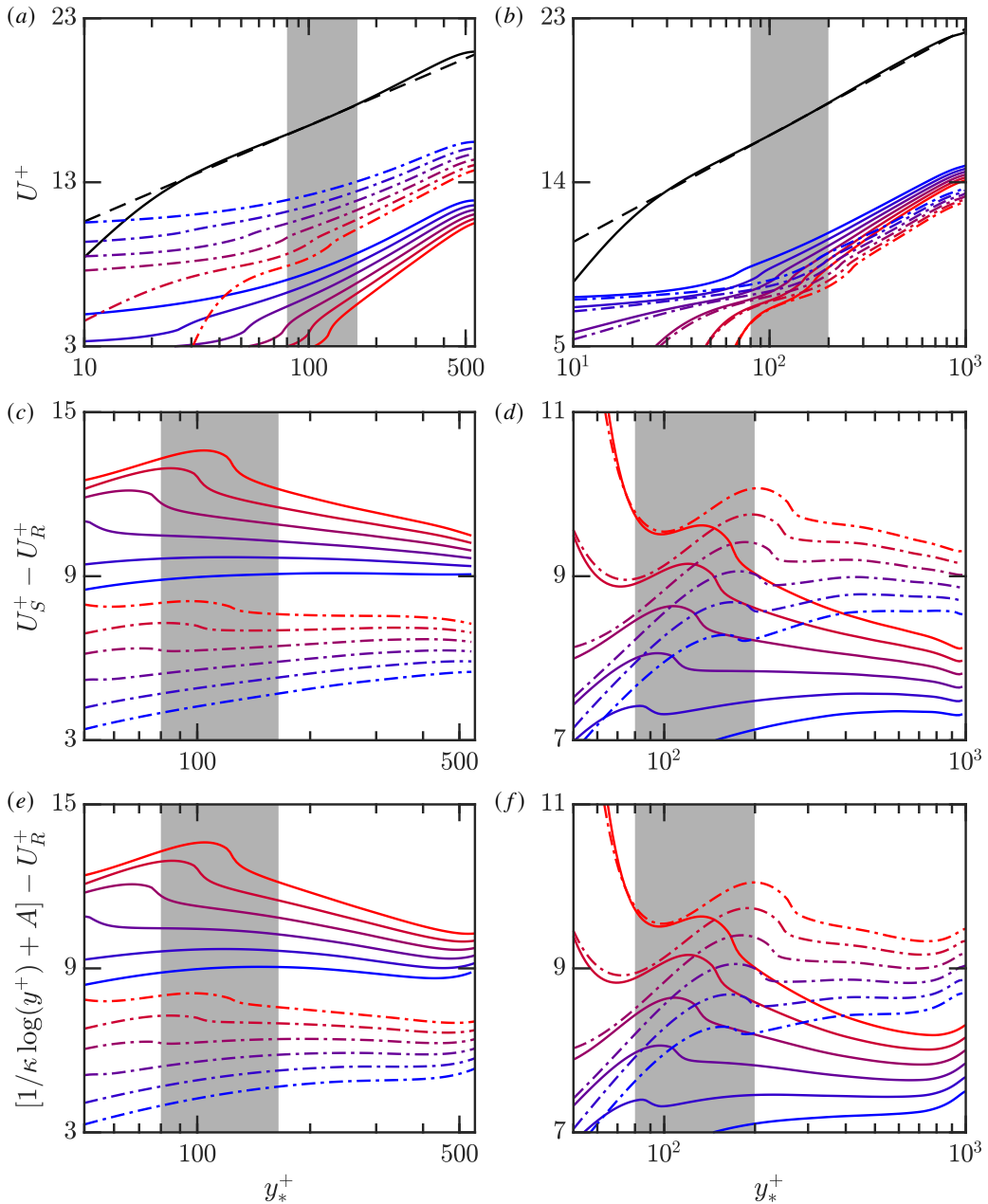

\cite{hama1954boundary} and \cite{clauser1956turbulent} noted that roughness generally leads to a downward but otherwise parallel shift, $\Delta U^+$, in the mean-velocity profile. It stems from this that $\Delta y$ and $u_\tau^*$ can be obtained by matching the shape of the mean-velocity profile over roughness, $U^+$, to a $\log(y_*^+)$, assuming the latter accurately represents the corresponding smooth-wall profile. This matching is usually done iteratively. Figures \ref{fig:rough_delU}$(e,f)$ illustrate how, for the flows over our canopies, $U^+$ can be made logarithmic by selecting a suitable value of $\Delta y$ and taking $u_\tau^*$ based on the total shear stress at the zero-plane-displacement height. However, $U^+$ is not exactly logarithmic even in the logarithmic layer of a smooth-wall flow, so matching $U^+$ to a $\log(y_*^+)$ as in figures \ref{fig:rough_delU}$(e,f)$ is different from matching $U^+$ to the smooth-wall profile in the logarithmic layer as in figures \ref{fig:rough_delU}$(c,d)$. As an example, $\Delta y\approx0.1\delta-0.15\delta=0.5h-0.75h$ enforces a logarithmic mean-velocity profile for case C144$_{550}$, but with a non-smooth-wall-like $\kappa_c$, as evidenced in figures \ref{fig:rough_delU}$(a,e)$. In addition, the mean-velocity profile above the logarithmic layer is different from a corresponding smooth-wall profile, suggesting a breakdown of outer-layer similarity, even though $U^+$ was made logarithmic. Alternatively, by imposing $\Delta y\approx0.1\delta=0.5h$, we may recover a smooth-wall-like logarithmic layer, as depicted in figure \ref{fig:rough_delU}$(c)$, where $\kappa_c\approx\kappa_s\approx0.39$. Nevertheless, the outer-wake region is still not smooth-wall-like, which would still break full outer-layer similarity. Moreover, for the intrusive cases C432$_{550}$ and O800$_{1000}$, where the near-wall turbulence is completely disrupted by the element-induced flow, the mean-velocity profile could still be enforced to take a logarithmic or smooth-wall-like shape within the 'logarithmic layer,' as shown in figures \ref{fig:rough_delU}$(c-f)$. Nevertheless, the flow above this 'logarithmic layer' is never smooth-wall-like. In contrast, outer-layer similarity may still be achieved without recovering a complete smooth-wall-like logarithmic region, as illustrated for case O400$_{1000}$ in figure \ref{fig:rough_delU}$(b,d,f)$, where the lower part of the logarithmic region is perturbed by the canopy but the flow above $y_*^+\approx130$ is essentially smooth-wall-like when $\Delta y =0.1\delta$. The above suggests that outer-layer similarity cannot be recovered simply by artificially matching $U^+$ to a $\log(y_*^+)$, or a smooth-wall profile, exclusively in the logarithmic layer. The matching should be for any height above the roughness sublayer.

\begin{figure}
    \vspace*{2mm}
    \centering
    \includegraphics[width=\textwidth]{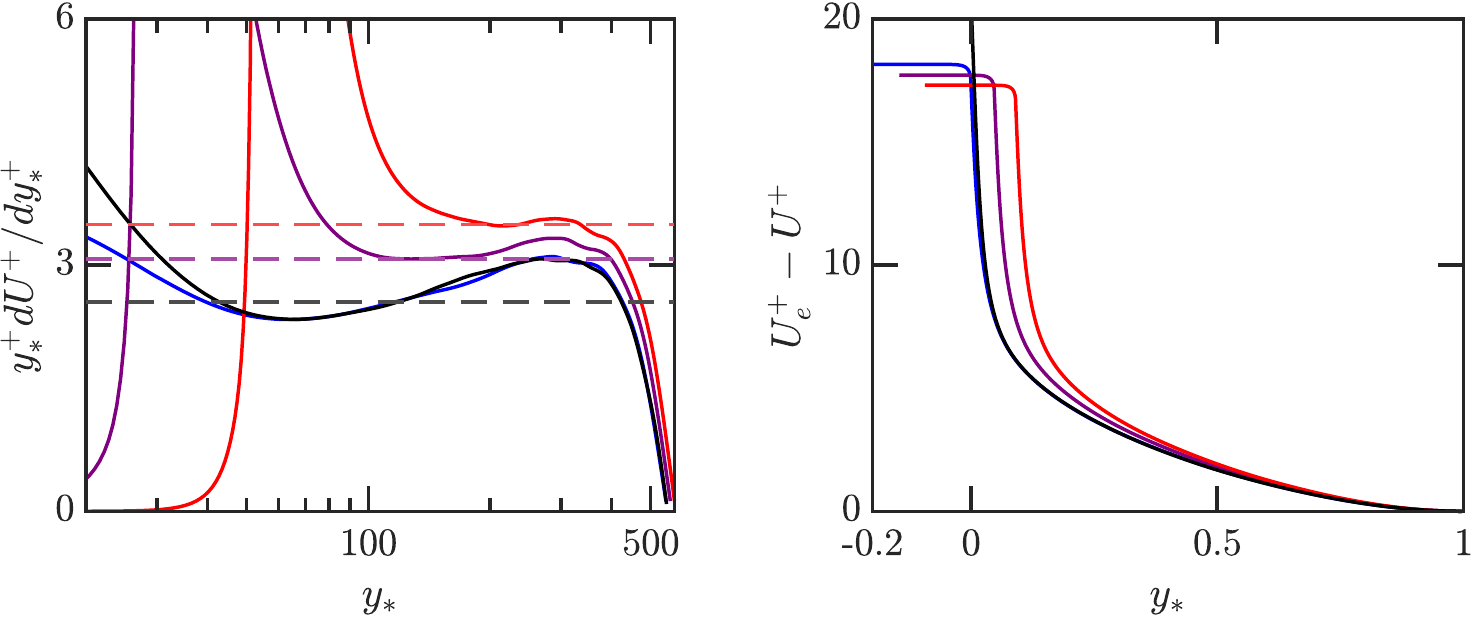}
    \put (-137mm,54mm) {$(a)$}
    \put (-65mm,54mm) {$(b)$}
    \caption{$(a)$ Diagnostic function and $(b)$ defect-law velocity profile, scaled with $y_*$ and $u_\tau^*$ for case C36$_{550}$. \protect\solidk, reference smooth-wall profile C$_{550}$; \protect\solidb, \protect\solidv, \protect\solidr, canopy statistics based on $\Delta y=0$, $0.05\delta$, and $0.1\delta$, respectively; and \protect\dashedk, \protect\dashedv, \protect\dashedr values of $1/\kappa$ for smooth-wall and canopy flow, where $\kappa_s\approx0.39$ and $\kappa_c\approx0.32$ and 0.28, respectively.}
    \label{fig:rough_diag}
    \vspace*{5mm}
    \centering
    \includegraphics[width=\textwidth]{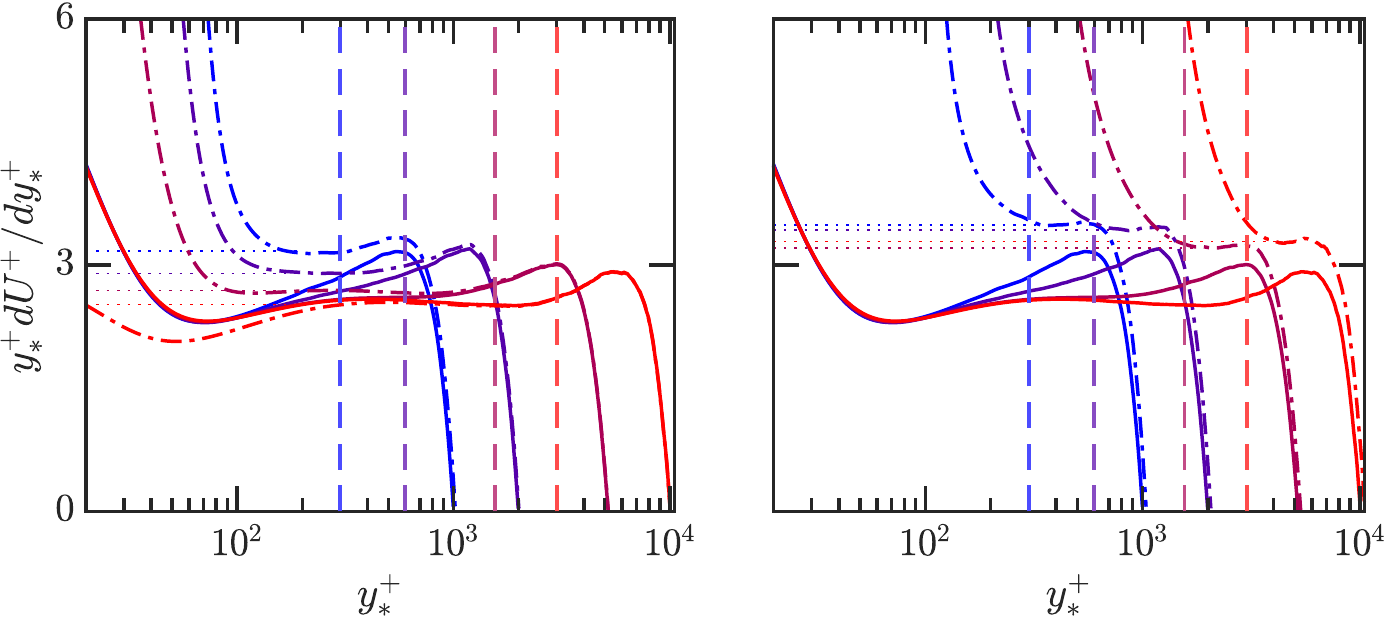}
    \put (-137mm,57mm) {$(a)$}
    \put (-65mm,57mm) {$(b)$}
    \caption{Diagnostic function for smooth-wall channel flows at $Re_\tau$ increasing from blue to red: $Re_\tau\approx1000, 2000, 5000$, and $10000$. The solid lines represent $y^+dU^+/dy^+$, with $\Delta y=0$; the dash-dotted lines in $(a)$ represent $\beta$ with $\Delta y^+\approx40, 27, 13, 7$, so that a plateau is enforced within the logarithmic layer as in \cite{manes2011turbulent} and \cite{okazaki2021describing, okazaki2022turbulent}; the dash-dotted lines in $(b)$ represent $\beta$ with $\Delta y^+\approx80, 139, 311, 1004$, so that the extent of plateau is maximised as in \cite{breugem2006influence}, \cite{suga2010effects}, \cite{kuwata2017direct}, \cite{rosti2017numerical}, \cite{fang2018influence}, \cite{shen2020direct}, \cite{kazemifar2021effect}, \cite{okazaki2021describing, okazaki2022turbulent}, and \cite{karra2022pore}. Notice that \cite{okazaki2021describing, okazaki2022turbulent} use both methods (see figures 5$(c)$ and 5$(a,b,d)$ in \cite{okazaki2021describing} and figures 12 and 21 in \cite{okazaki2022turbulent}). With $Re_\tau$ increasing, the horizontal dotted lines correspond to $\kappa_c\approx(0.32,0.34,0.37,0.40)$ in $(a)$ and $\kappa_c\approx(0.29,0.29,0.31,0.30)$ in $(b)$, and the vertical lines represent the upper bound of the logarithmic layer, $y/\delta\approx0.3$. $Re_\tau\approx1000,5000$ data are from \cite{lee2015direct}, $Re_\tau\approx2000$ data is from \cite{hoyas2006scaling}, and $Re_\tau\approx10000$ data is from \cite{hoyas2022wall}.}
    \label{fig:smooth_diag}
\end{figure}

Within the logarithmic layer, enforcing $U^+$ to be logarithmic, or smooth-wall-like, is essentially equivalent to enforcing $\beta$ to have a plateau, or a smooth-wall-like region. While both the shape of $U^+(y_*^+)$ and $\beta(y_*^+)$ contain the same information, $\beta$ is more sensitive to deviations from the smooth-wall reference profile, and it directly portrays the value of $1/\kappa$ if it exhibits a plateau in the logarithmic layer. For these reasons, some recent studies rely on $\beta$ to predict $\Delta y$ and $u_\tau^*$ \citep{mizuno2011mean, kuwata2017direct, okazaki2021describing, okazaki2022turbulent}. \cite{breugem2006influence} and \cite{suga2010effects} argued that the slope of $U^+$ v.s. $\log(y_*^+)$ must be constant within the logarithmic layer, and the profile of $(y+\Delta y)dU/dy$ should therefore exhibit a plateau with value $u_\tau^*/\kappa$. It is worth mentioning that although the method and argument by \cite{breugem2006influence} and \cite{suga2010effects} have been adapted by some recent studies, these studies actually maximise the extent of the plateau in the entire flow, instead of just within the logarithmic layer (e.g. figure 5 in \cite{breugem2006influence}, figure 6 in \cite{rosti2017numerical}, and figure 8 in \cite{shen2020direct}). \colcomm{However, let us illustrate using one of our present cases that there are instances when enforcing a plateau within the logarithmic layer or maximising the extent of the plateau may result in a breakdown of outer-layer similarity. Figure \ref{fig:rough_diag}$(a)$} shows the diagnostic function of case C36$_{550}$, which consists of closely packed elements. A plateau in $\beta$ emerges within the logarithmic layer when picking $\Delta y=0.05\delta=0.25h$, and the extent of this plateau maximises when picking $\Delta y=0.1\delta=0.5h$. The resulting Kármán constant, $\kappa_c\approx0.32$ or $0.29$ respectively, is much smaller than the smooth-wall value, $\kappa_s\approx0.39$, implying a significant modification of the outer-layer mean-velocity profile by the substrate. \colcomm{For dense} canopies like C36$_{550}$, however, we would expect the overlying turbulent flow to 'skim' over the canopy tips, as the turbulent eddies are precluded from penetrating within and interacting with the full height of the canopy, to the point that the flow above the tips resembles a smooth-wall flow \citep{sharma2020turbulent_b, brunet2020turbulent}. Figure \ref{fig:rough_diag}$(b)$ illustrates that the scaling based on $\Delta y=0$ produces a smooth-wall-like diagnostic function for case C36$_{550}$, suggesting that the overlying flow indeed perceives an origin in the vicinity of the canopy-tip plane. The underlying problem is that the emergence of a logarithmic layer relies on a large Reynolds number, $\delta^+$, such that the only available length scale in the overlap region, $\nu/u_\tau\ll y\ll\delta$, is the wall-normal distance, $y$ \citep{townsend1976structure}. Within this overlap region, only a dimensionless constant, $\kappa=u_\tau/(y\partial U/\partial y)$, can be constructed \citep{luchini2017universality}. However, if $\delta^+$ is not sufficiently  large, $u_\tau/(y\partial U/\partial y)$ need not exhibit a flat plateau, as shown in figure \ref{fig:smooth_diag}. \colcomm{For} smooth-wall flows, even $Re_\tau\approx5000$ is not yet sufficient for the diagnostic function to exhibit a completely flat plateau, \colcomm{according to numerical evidence} \citep{lee2015direct, hoyas2022wall}. The local value of \colcomm{$\beta\approx1/\kappa$} in the logarithmic layer exhibits a dependence on $y/\delta$, or $y^+/Re_\tau$, caused by the contamination from the wake above \citep{jimenez2007we, mizuno2011mean, luchini2018structure}. As shown in figure \ref{fig:smooth_diag}, even for smooth-wall flows, so long as $Re_\tau\lesssim5000$, enforcing a plateau in $\beta$ would overlook this dependence, and result in nonphysical zero-plane displacements \colcomm{and values of $\kappa$ down to 0.3, as those listed in the figure. The above results for smooth walls suggest that} artificially prescribing a plateau in $\beta$, while not enforcing the similarity in the wake region, could result in an apparent but false breakdown of outer-layer similarity and values for $\kappa$ that are consistently lower than the true smooth-wall value, as those listed in table \ref{tab:studies}. It is worth mentioning that the logarithmic layer of a smooth-wall channel is generally understood to span from $y^+\approx80$ to $y/\delta\approx0.3$. It is therefore not possible to define $\kappa$ meaningfully for flows at Reynolds numbers $\delta^+\lesssim300$, for which there is no significant range of $y$ in which a logarithmic layer can manifest. Note that some of the flows in table \ref{tab:studies} fall in this low-$Re$ range.

\colcomm{To address the above issues, we propose to} obtain $u_\tau^*$ and $\Delta y$ by minimising the deviation between the smooth-wall and canopy diagnostic function everywhere above the roughness sublayer, not just in the logarithmic layer. It will be demonstrated in \S \ref{subsec:Sens} and \S \ref{subsec:Univ} that this method consistently recovers outer-layer similarity.

\subsection{Sensitivity of canopy diagnostic function}\label{subsec:Sens}
Outer-layer similarity can only be expected to appear above the roughness sublayer, of height $y_{r}$, above which the canopy diagnostic function, $\beta_c$, should be smooth-wall-like. Because the extent of roughness effects can vary depending on the canopy density, as shown in figure \ref{fig:cond_stats}, $y_{r}$ needs to be determined separately for each canopy \citep{jimenez2004turbulent, brunet2020turbulent}. Care must be taken, because if part of the roughness sublayer is included in the region where outer-layer similarity is sought, the values of $\Delta y$ and $u_\tau$ may be distorted. Therefore, the latter region needs to be sufficiently far away from the wall, $y>y_{r}$, such that all surface effects have vanished. We determine the values of $\Delta y$ and $u_\tau$ that recover a smooth-wall-like diagnostic function, or mean velocity profile, as those that minimise the deviation between $\beta_s$ and $\beta_c$, the smooth-wall and canopy diagnostic functions, above $y_{r}$.  By this method,  we recover a smooth-wall-like $\beta_c$ in the outer layer, including both the logarithmic layer and the 'wake' region.

\begin{figure}
    \vspace*{3mm}
    \centering
    \includegraphics[width=\textwidth]{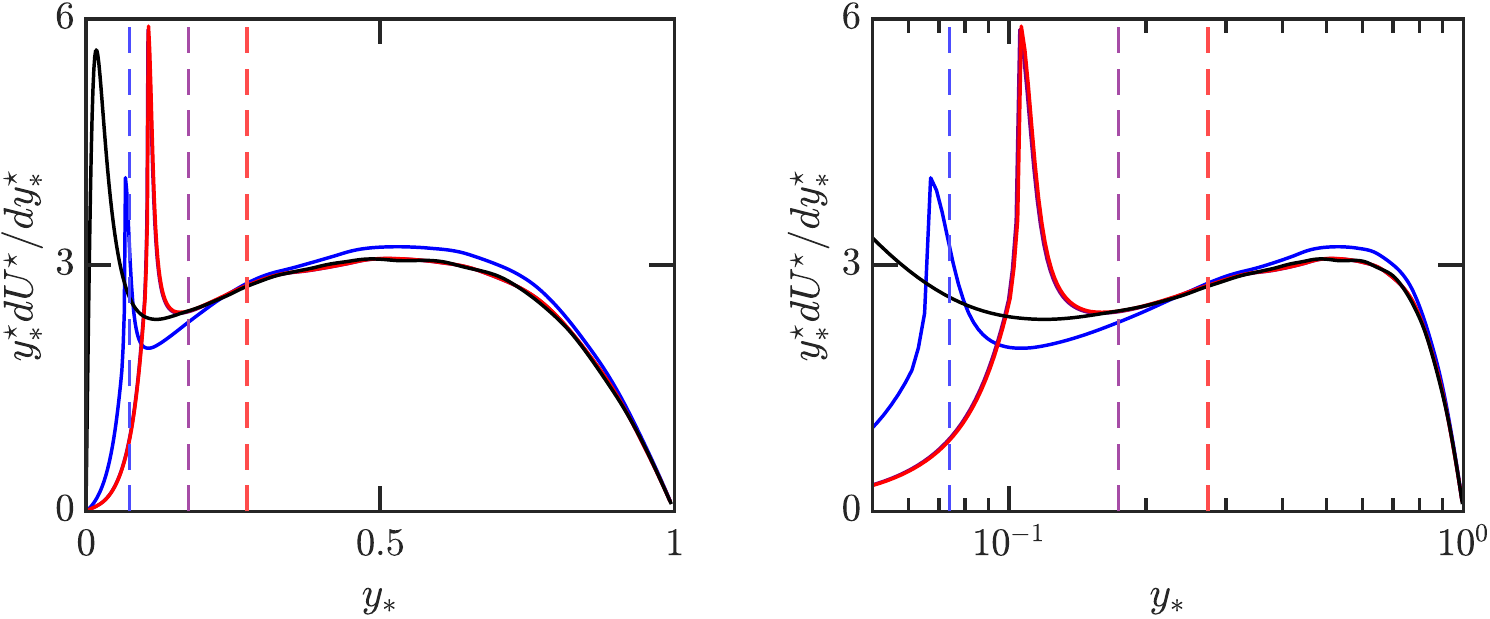}
    \put (-137mm,54mm) {$(a)$}
    \put (-65mm,54mm) {$(b)$}
    \caption{Diagnostic function of case C144$_{550}$ scaled with $\Delta y$ and $u_\tau^\star$, the friction velocity decoupled from $y_*=0$, that minimise the r.m.s. deviation from the smooth-wall profile above $y_{r}$. \protect\solidk, reference smooth-wall profile at $Re_\tau\approx550$; \protect\solidb, \protect\solidv, \protect\solidr, canopy statistics based on \protect\dashedb $y_{r}=0.074\delta$, \protect\dashedv $y_{r}=0.174\delta$, and \protect\dashedr $y_{r}=0.274\delta$, respectively.}
    \label{fig:I144_diag}

    \vspace*{5mm}

    \centering
    \includegraphics[width=\textwidth]{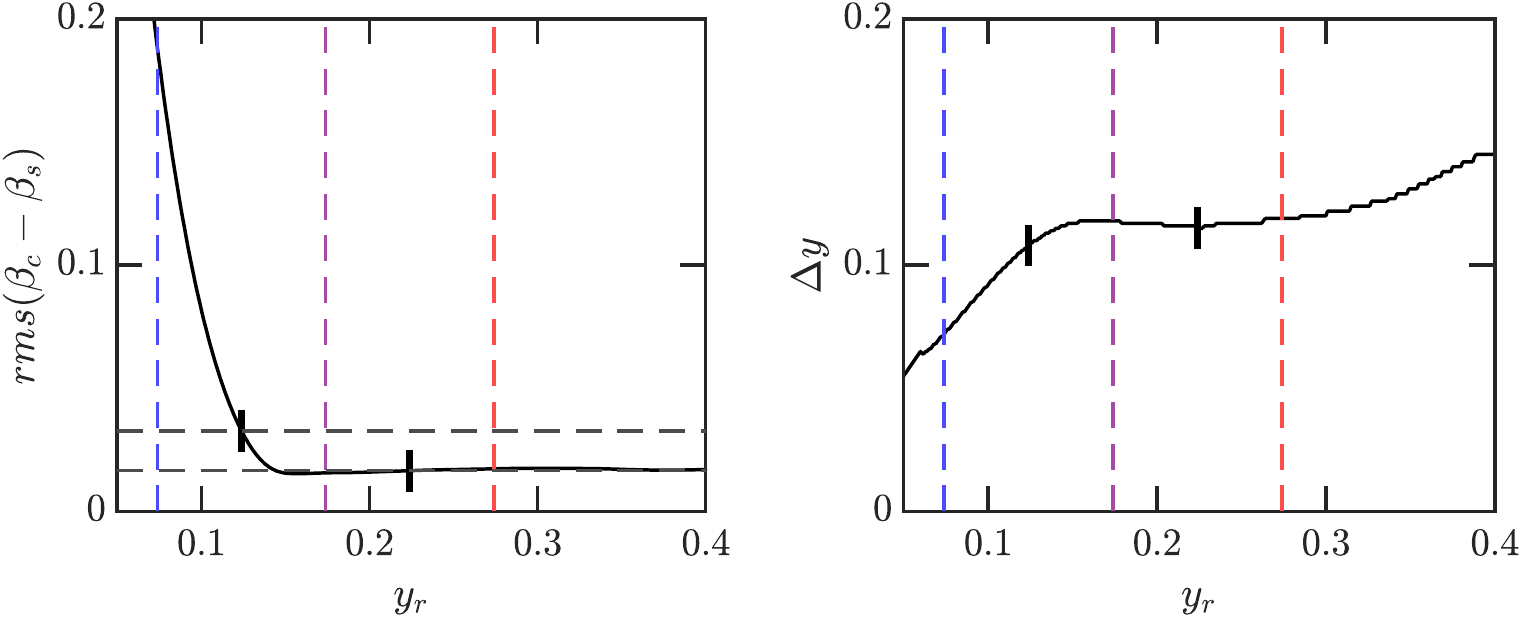}
    \put (-137mm,53mm) {$(a)$}
    \put (-65mm,53mm) {$(b)$}
    \caption{$(a)$ R.m.s. deviation between $\beta_s$ and $\beta_c$ above $y_{r}$ and $(b)$ zero-plane displacement, $\Delta y$, versus $y_{r}$ for case C144$_{550}$. The horizontal dashed lines mark once and twice the baseline error; and the thick black vertical markers denote the lower and upper bounds for $y_{r}$ above which outer-layer similarity recovers. The vertical dashed lines are as in figure \ref{fig:I144_diag}.}
    \label{fig:beta_err_dy}
\end{figure}

As an example, for case C144$_{550}$, figure \ref{fig:I144_diag} portrays $\beta_c$ scaled with $y_*$ and $u_\tau^\star$, the friction velocity decoupled from $y_*=0$, based on three tentative values of $y_{r}$. For $y_{r}=0.074\delta=0.37h=0.28s$, $\beta_c$ is not smooth-wall-like, even if the deviation between $\beta_c$ and $\beta_s$ above $y_{r}$ is minimised. This is because the flow at this $y_{r}$ is still directly perturbed by the canopy elements, as $y_{r}$ is too small compared to $s$, which provides an estimate of the roughness sublayer thickness, as evidenced in figure \ref{fig:cond_stats}. In turn, adopting $y_{r}=0.274\delta=1.37h=1.04s$ overestimates the roughness sublayer thickness, as $\beta_c$ is already smooth-wall-like based on $y_{r}\gtrsim0.174\delta=0.87h=0.66s$, implying that the height affected by the canopy in this case is lower than $y_r/s\approx2-3$ reported in \cite{abderrahaman2019modulation} and \cite{sharma2020turbulent_b}. For this canopy, we can conclude that outer-layer similarity is recovered when $y_{r}\approx0.66s$, and beyond this limit $\beta_c$ and $\Delta y$ are insensitive to the lower bound from which outer-layer similarity is enforced, as illustrated in figure \ref{fig:I144_diag} and figure \ref{fig:beta_err_dy}$(b)$. The r.m.s. deviation between $\beta_c$ and $\beta_s$ above $y_{r}$ is minimised if $y_{r}$ is sufficiently far away from the canopy, as shown in figure \ref{fig:beta_err_dy}$(a)$.

In the above, we have let $\kappa_c=\kappa_s$ and $u_\tau^\star$ be independent of $\Delta y$, and we obtain the values of $\Delta y$ and $u_\tau^\star$ for each $y_r$ by minimising the difference between $\beta_c$ and $\beta_s$ for any $y$ above $y_r$. However, if $y_r$ is not carefully chosen, the corresponding $\Delta y$ and $u_\tau^\star$ may not result in a smooth-wall-like $\beta_c$ above $y_r$, as illustrated in figure \ref{fig:I144_diag}. Therefore, we need to identify an appropriate lower bound for $y_r$ to correctly assess outer-layer similarity, sufficiently far away from the wall for all surface effects to vanish. Starting at too low values, as $y_r$ increases, the r.m.s. deviation of $\beta$ in figure \ref{fig:beta_err_dy}$(a)$ decreases and eventually stabilises at a baseline-error level, which may be used to guide the selection of the lower bound of $y_r$. Here, we propose a lower bound of $y_{r}$ such that the r.m.s. error is twice the baseline error. Above this lower bound, the deviation between $\beta_c$ and $\beta_s$ is small, suggesting that $\beta_c$ scaled with $y_*$ and $u_\tau^\star$ is essentially smooth-wall-like. On the other hand, $y_{r}$ should not be adopted excessively far above the surface, because outer-layer similarity cannot be properly examined on too small a portion of the flow. We propose an upper bound for $y_{r}$ that is $0.1\delta$ above the lower bound, as the r.m.s. of $\beta_c-\beta_s$ varies little beyond this.

\begin{figure}
    \vspace*{-1mm}
    \centering
    \hspace*{-3mm}\includegraphics[width=.95\textwidth]{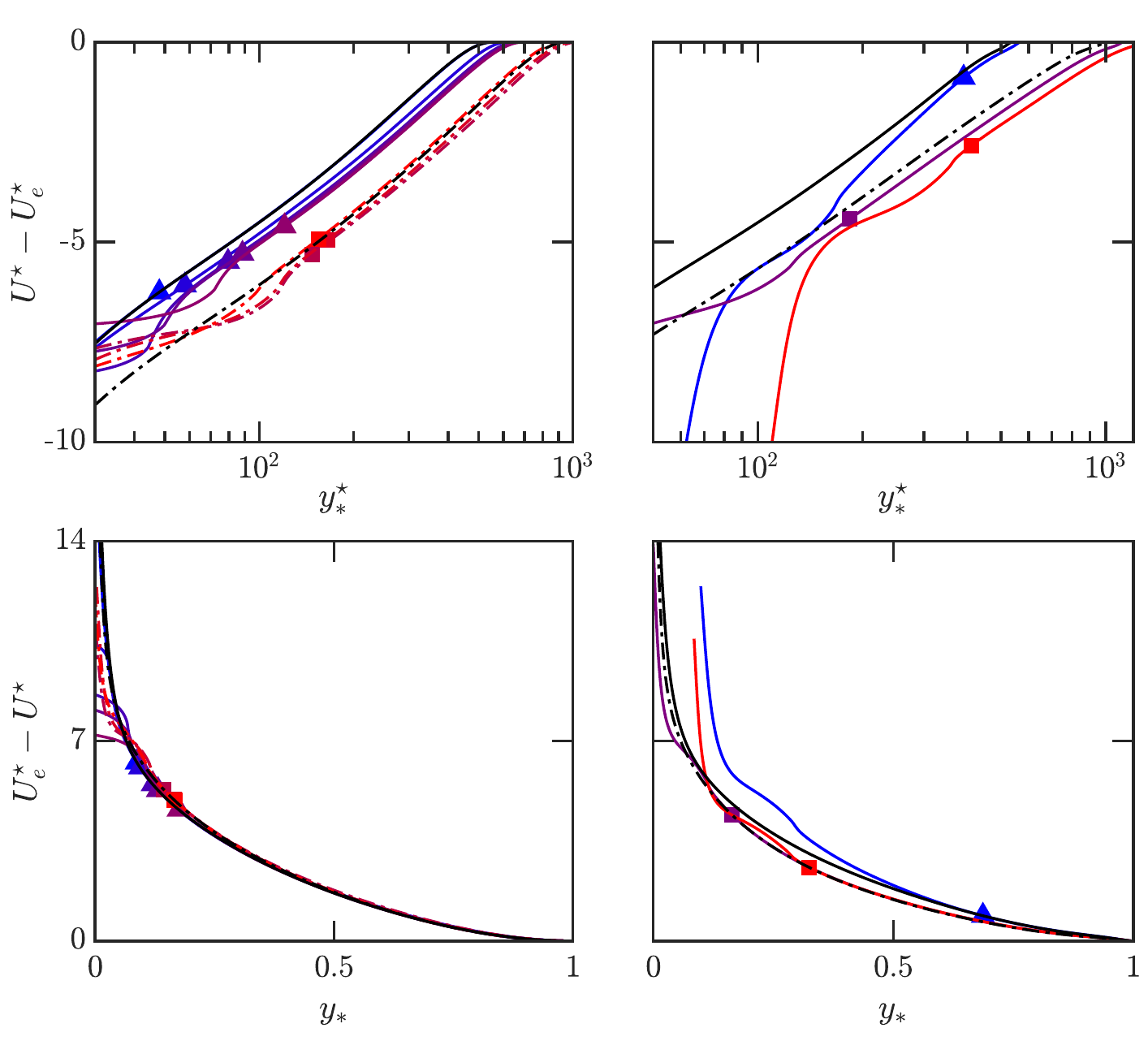}
    \put (-128mm,111mm){$(a)$}
    \put (-61mm,111mm) {$(b)$}
    \put (-128mm,54mm) {$(c)$}
    \put (-61mm,54mm) {$(d)$}
    \vspace*{-2mm}
    \caption{Mean-velocity profiles of the canopy and smooth-wall flows. $(a,c)$, full-channel cases with less intrusive canopies, C36$_{550}$ to C144$_{550}$ and C216$_{900}$ to C432$_{900}$; $(b,d)$, open-channel cases. The coloured solid and dash-dotted lines represent cases as in in figure \ref{fig:cond_stats}. The black solid lines represent smooth-wall reference profiles at $Re_\tau\approx900-1000$, and the black dashed lines represent smooth profiles at $Re_\tau\approx550$. The square markers represent the lower bound of $y_r$ for cases at $Re_\tau\approx550$, and the triangle markers represent that for those at $Re_\tau\approx900-1000$.}
    \label{fig:df}
\end{figure}
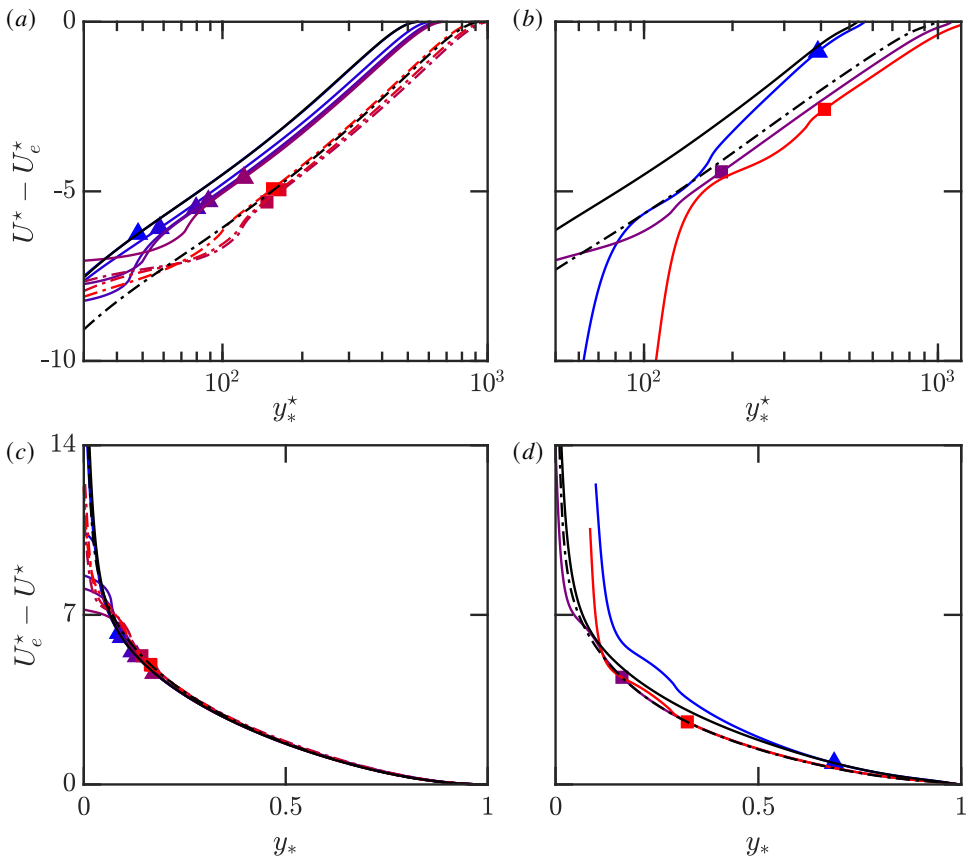
\vspace*{-1mm}

Using the method we propose, $y_{r}$ results in $\Delta y$ and $u_\tau^\star$ that consistently recover outer-layer similarity, with the resulting $\Delta y$ being insensitive to the particular choice of $y_r$ within the range proposed above, as evidenced by the flat region in Figure \ref{fig:beta_err_dy}$(b)$. As shown in figure \ref{fig:beta_err_dy}$(a)$, the height above which outer-layer similarity recovers for case C144$_{550}$ is $y_{r}=(0.174\pm0.05)\delta=(0.87\pm0.25)h=(0.66\pm0.19)s$. This is smaller than the typical roughness sublayer thickness. which extends up to $2-3h$ above the tips \citep{jimenez2004turbulent, brunet2020turbulent}. Nevertheless, figure \ref{fig:cond_stats} illustrates that the dominant length scale for the intensity of the texture-coherent flow is the element spacing, and these texture-coherent flows essentially vanish at one spacing above the tips. For intermediate to dense cases with $\lambda\gtrsim0.1$, where $s/h\lesssim1$, we observe the recovery of outer-layer similarity for $y_r\approx s\lesssim h$. Additionally, because $y_{r}/\delta$ is smaller than the typical upper bound of a logarithmic layer, $y\approx0.3\delta$, we expect the recovery of a smooth-wall-like logarithmic layer. Based on the confidence interval for $y_{r}$, we obtain the zero-plane displacement,  $\Delta y=(0.113\pm0.005)\delta=(0.57\pm0.03)h$, as shown in figure \ref{fig:beta_err_dy}$(b)$, indicating that the outer-layer flow perceives an origin $y_*=0$ at a depth roughly half-way between the floor and the canopy tips. However, the height of $u_{\tau}^\star$, corresponding to a smooth-wall-like $\kappa$, is $y_{u_{\tau}^\star}=(0.553\pm0.036)\delta=(2.77\pm0.18)h$, which is below the canopy floor (note that $y_{u_{\tau}^\star}$ is decoupled from $y_*=0$).

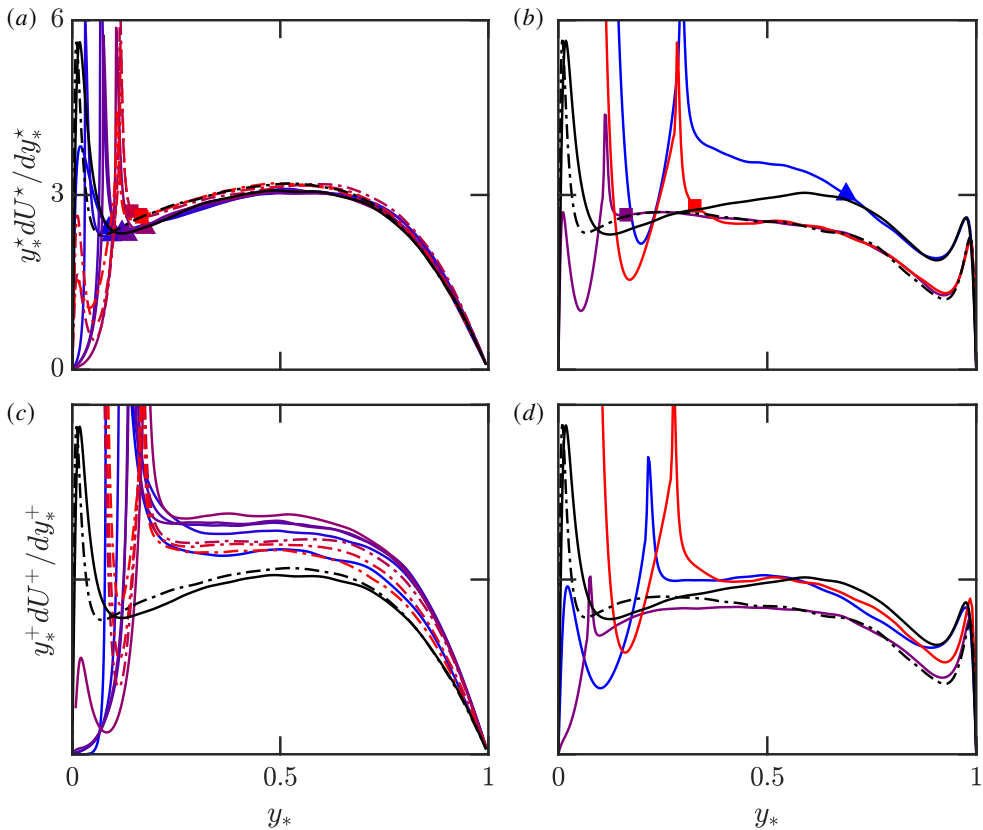
\begin{figure}
    \vspace*{2mm}
    \centering
    \hspace*{-3mm}
    \includegraphics[width=.95\textwidth]{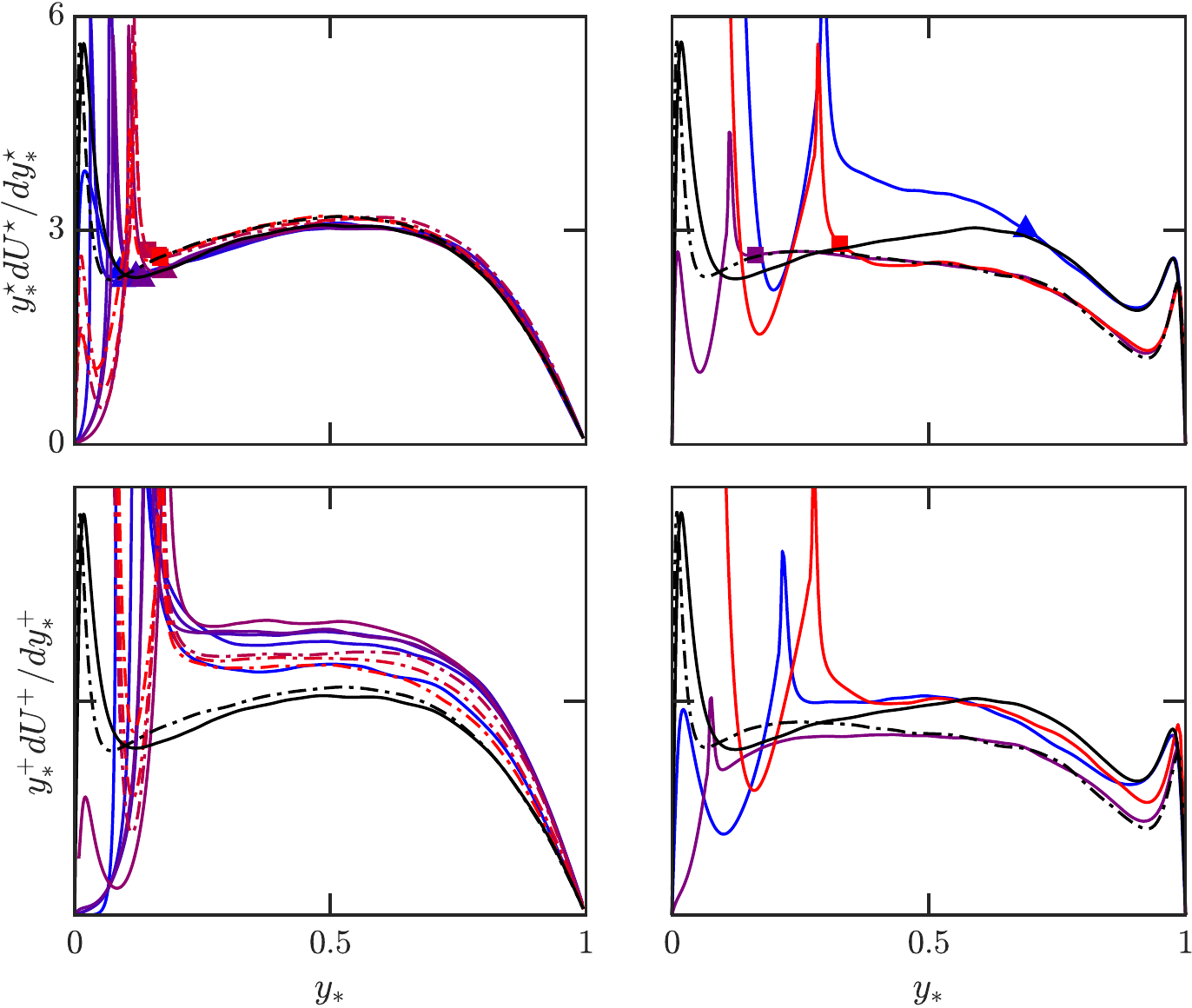}
    \put (-129mm,106mm){$(a)$}
    \put (-62mm,106mm) {$(b)$}
    \put (-129mm,54mm) {$(c)$}
    \put (-62mm,54mm) {$(d)$}
    \caption{Diagnostic functions of the canopy and smooth-wall flows. $(a,c)$, full-channel cases; $(b,d)$, open-channel cases. $(a,b)$ enforce smooth-wall-like $\beta_c$ above $y_{r}$; $(b,d)$ enforce plateaus in $\beta_c$. The lines and markers are as in figure \ref{fig:df}.}
    \label{fig:diag}
\end{figure}

The fact that outer-layer similarity can only be assessed above a minimum height $y_r$ implies that such similarity cannot be assessed for certain flows/substrates. In figures \ref{fig:df}$(c,d)$ and \ref{fig:diag}$(c)$, we observe a breakdown of outer-layer similarity for the open-channel cases O400$_{550}$ ($y_{r}\geq0.636\delta$) and O800$_{1000}$ ($y_{r}\geq0.274\delta$). This is because these intrusive canopies perturb the overlying flow extensively, well into the region where a logarithmic layer would otherwise have been found. In contrast, for intermediate to dense cases, C36$_{550}$ to C144$_{550}$, C216$_{900}$ to C432$_{900}$, and O400$_{1000}$, which have less intrusive canopies, the canopy flow becomes smooth-wall-like no higher than $y_{r}=0.224\delta$, allowing for the recovery of a smooth-wall-like outer layer. The canopy diagnostic functions in figure \ref{fig:diag}$(a,b)$ also show that whether outer-layer similarity recovers depends on the extent of the roughness layer, which is associated with $y_{r}$. In contrast, prescribing a constant slope for $U^+$ v.s. $\log(y_*)$, or enforcing a plateau in $\beta_c$, can consistently result in a breakdown of outer-layer similarity and a Kármán constant different from the smooth-wall one, as illustrated in figures \ref{fig:diag}$(c,d)$.

\subsection{Universality or non-universality of the Kármán constant}\label{subsec:Univ}
In the above, the friction velocity, $u_\tau^\star$, was not computed from the stress at the zero-plane-displacement height, $y_*=0$, and was instead set as an independent variable, such that $\beta_c$ was smooth-wall-like and $\kappa$ was universal, $\kappa_c=\kappa_s$. Alternatively, the friction velocity, $u_\tau^*$, could be computed from the stress at $y_*=0$, and $\kappa_c$ found by enforcing a match of the diagnostic function to that of a smooth-wall profile. In the logarithmic layer, equation (\ref{eq:expect diag}) can be expressed as

\begin{figure}
    \vspace*{0mm}
    \centering
    \includegraphics[width=\textwidth]{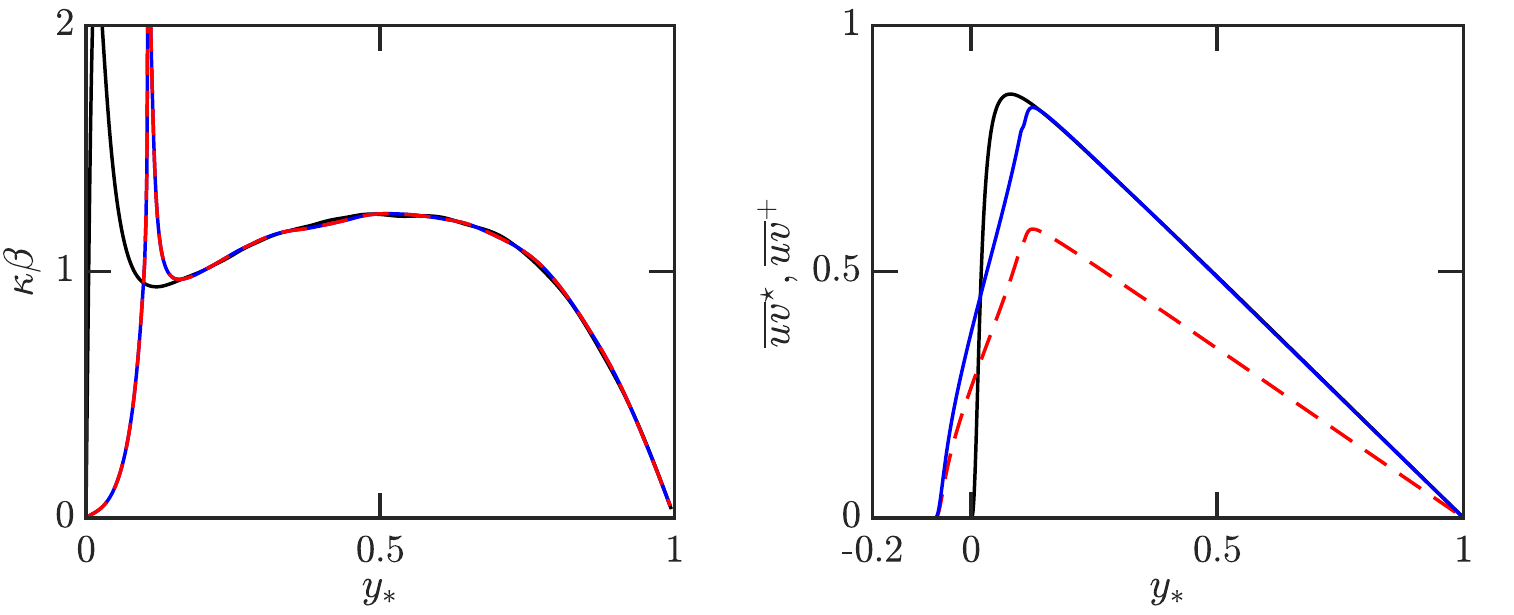}
    \put (-137mm,52mm) {$(a)$}
    \put (-65mm,52mm) {$(b)$}
    \vspace*{-1mm}
    \caption{$(a)$ Modified diagnostic function, $\kappa\beta$, and $(b)$ Reynolds shear stress of case C144$_{550}$. \protect\solidk, reference smooth-wall profiles at $Re_\tau\approx550$; \protect\dashedr, canopy profile scaled with $u_\tau^\star$ and $y_*$; \protect\solidb, canopy profile scaled with $u_\tau^*$ and $y_*$.}
    \label{fig:kappa_beta}
\end{figure}

\begin{figure}
    \centering
    \includegraphics[width=\textwidth]{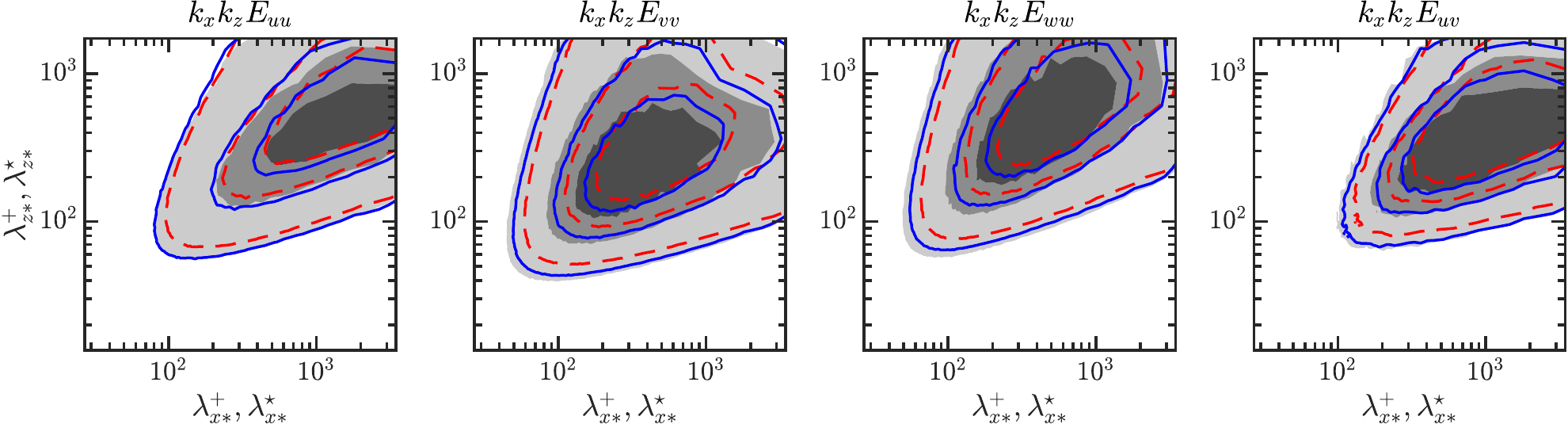}
    \put (-137mm,34mm) {$(a)$}
    \put (-100mm,34mm) {$(b)$}
    \put (-66mm,34mm) {$(c)$}
    \put (-33mm,34mm) {$(d)$}
    \caption{Pre-multiplied spectral energy densities for case C144$_{550}$ (line contours) and reference smooth-wall case C$_{550}$ (filled contour) normalised by their respectively r.m.s. level at $y_*=0.3\delta$. \protect\dashedr, canopy statistics scaled with $u_\tau^\star$ and $\nu$; \protect\solidb, canopy statistics scaled with $u_\tau^*$ and $\nu$. The contours are at 0.01, 0.05 and 0.1 times the r.m.s. level.}
    \label{fig:sp_I144}
\vspace*{-8mm}
\end{figure}
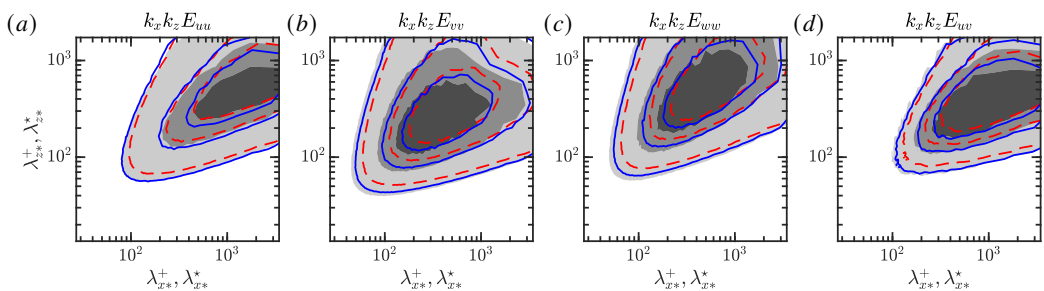

\begin{align}
y_*\frac{dU}{dy_*} \frac{\kappa_s}{u_\tau^\star}=\kappa_s\beta(y_*,u_\tau^\star)&\approx1,
\label{eq:modified diag1}\\
y_*\frac{dU}{dy_*} \frac{\kappa_c}{u_\tau^*}=\kappa_c\beta(y_*,u_\tau^*)&\approx1,
\label{eq:modified diag2}
\end{align}

\noindent from which we can observe that the group $\kappa/u_\tau$ needs to have a certain value, but the mean-velocity profile does not provide information on whether this should be achieved by fixing $\kappa$ and finding $u_\tau$, or vice versa, or else. As long as $\kappa_c/u_\tau^* = \kappa_s/u_\tau^\star$, the modified logarithmic profile in equation (\ref{eq:modified diag2}) is recovered. Figure \ref{fig:kappa_beta}$(a)$ shows that whether $u_\tau^\star$ or $\kappa_c\neq\kappa_s\approx0.4$ is set as an independent parameter, the resulting mean-velocity profile or diagnostic function  is still smooth-wall-like. In this section, we discuss the implications of fixing either $u_\tau$ or $\kappa$. While it is not possible to identify which option provides a complete outer-layer similarity by inspecting just the mean-velocity profile, other turbulent statistics collapse with the smooth-wall data when the friction velocity is evaluated at the zero-plane-displacement height. As an example, the Reynolds shear stress profile in figure \ref{fig:kappa_beta}$(b)$ collapses to smooth-wall data when scaled with $u_\tau^*$, compared to $u_\tau^\star$. If, alternatively, a universal $\kappa_c$ is enforced, the Reynolds shear stress would differ from smooth-wall values by a factor $(u_\tau^*/u_\tau^\star)^2$. The spectral density maps in figure \ref{fig:sp_I144} also show that outer-layer similarity is achieved with $u_\tau^*$ obtained from the shear at the zero-plane-displacement height, $y_*=0$, and $\kappa_c$ having a non-smooth-wall value.

\begin{figure}
    \centering
    \vspace*{3mm}
    \hspace*{-3mm}
    \includegraphics[width=.9\textwidth]{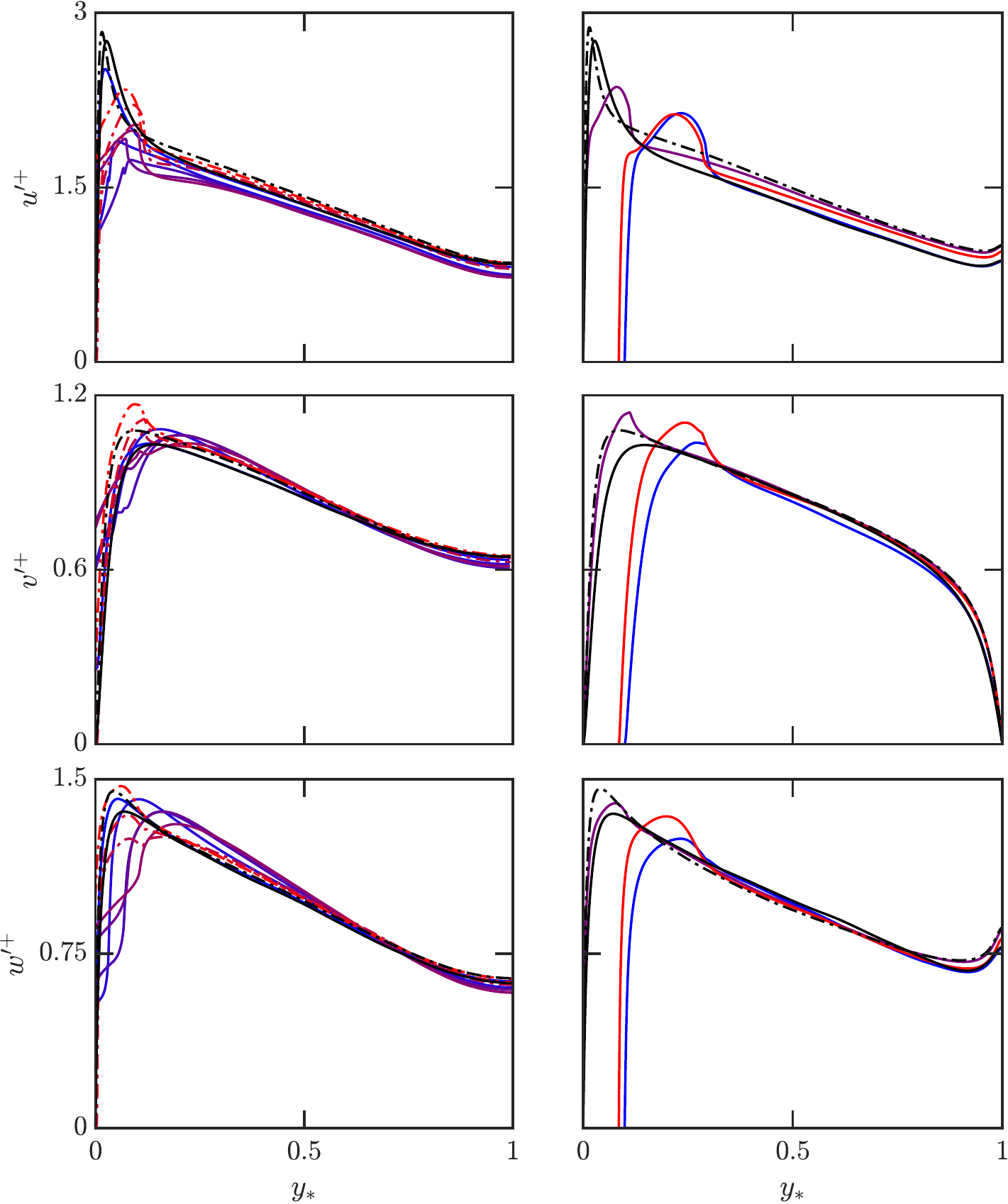}
    \put (-120mm,143mm) {$(a)$}
    \put (-57mm,143mm) {$(b)$}
    \put (-120mm,97mm){$(c)$}
    \put (-57mm,97mm) {$(d)$}
    \put (-120mm,51mm) {$(e)$}
    \put (-57mm,51mm) {$(f)$}
    \caption{R.m.s. velocity fluctuations scaled with $y_*$ and $u_\tau^*$. $(a,c,e)$, full-channel cases; $(b,d,f)$, open-channel cases. Lines are as in figure \ref{fig:df}.}
    \label{fig:rms}
\end{figure}

\begin{figure}
    \centering
    \vspace*{3mm}
    \includegraphics[width=\textwidth]{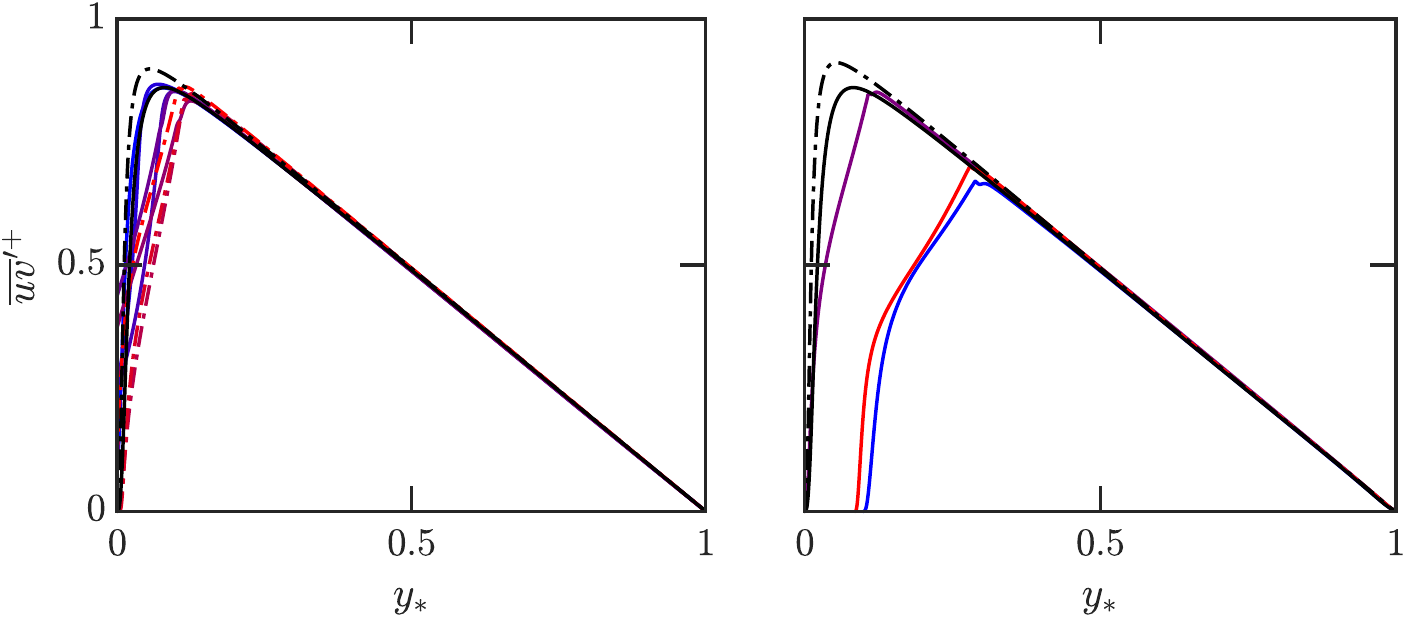}
    \put (-137mm,57mm) {$(a)$}
    \put (-65mm,57mm) {$(b)$}
    \caption{Reynolds shear stress scaled with $y_*$ and $u_\tau^*$. $(a)$, full-channel cases; $(b)$, open-channel cases. The lines are as in figure \ref{fig:df}.}
    \label{fig:uvm}

    \vspace{1cm}
    
    \includegraphics[width=\textwidth]{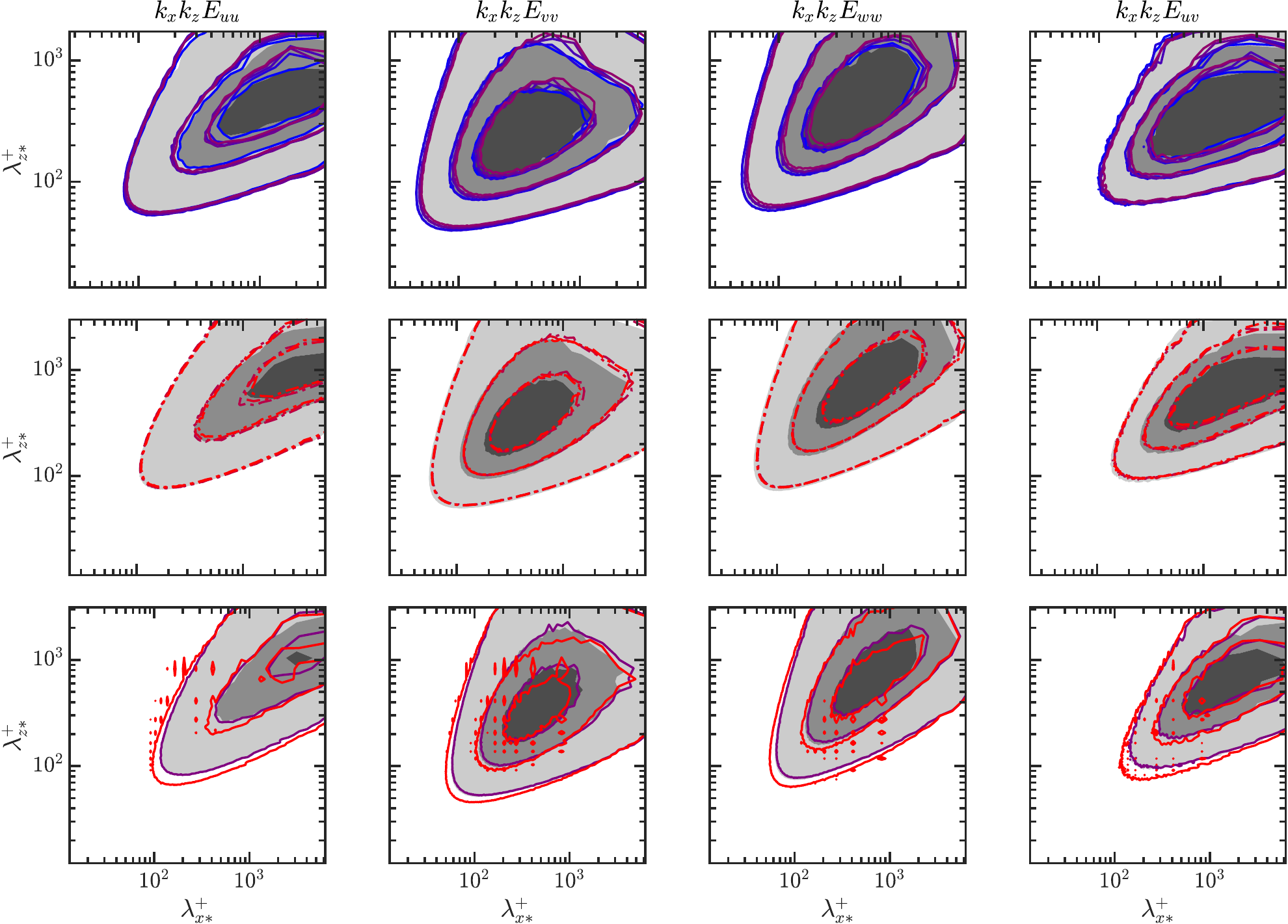}
    \put (-137mm,92mm) {$(a)$}
    \put (-100mm,92mm) {$(b)$}
    \put (-66mm,92mm) {$(c)$}
    \put (-33mm,92mm) {$(d)$}
    \put (-137mm,62mm) {$(e)$}
    \put (-100mm,62mm) {$(f)$}
    \put (-66mm,62mm) {$(g)$}
    \put (-33mm,62mm) {$(h)$}
    \put (-137mm,32mm) {$(i)$}
    \put (-100mm,32mm) {$(j)$}
    \put (-66mm,32mm) {$(k)$}
    \put (-33mm,32mm) {$(l)$}
    \caption{Pre-multiplied spectral energy densities for canopy-flow (line contours) and reference smooth-wall cases (filled contours) normalised by their respective r.m.s. level at $y_*=0.3\delta$. $(a-d)$ represent the full-channel cases; and $(e-h)$ and $(i-l)$ represent the open-channel cases. The line colour scheme is as in figure \ref{fig:df}. The contours are at 0.01, 0.05 and 0.1 times the mean-square fluctuation level.}
    \label{fig:sp_03}
\end{figure}

\begin{figure}
    \centering
    \includegraphics[width=\textwidth]{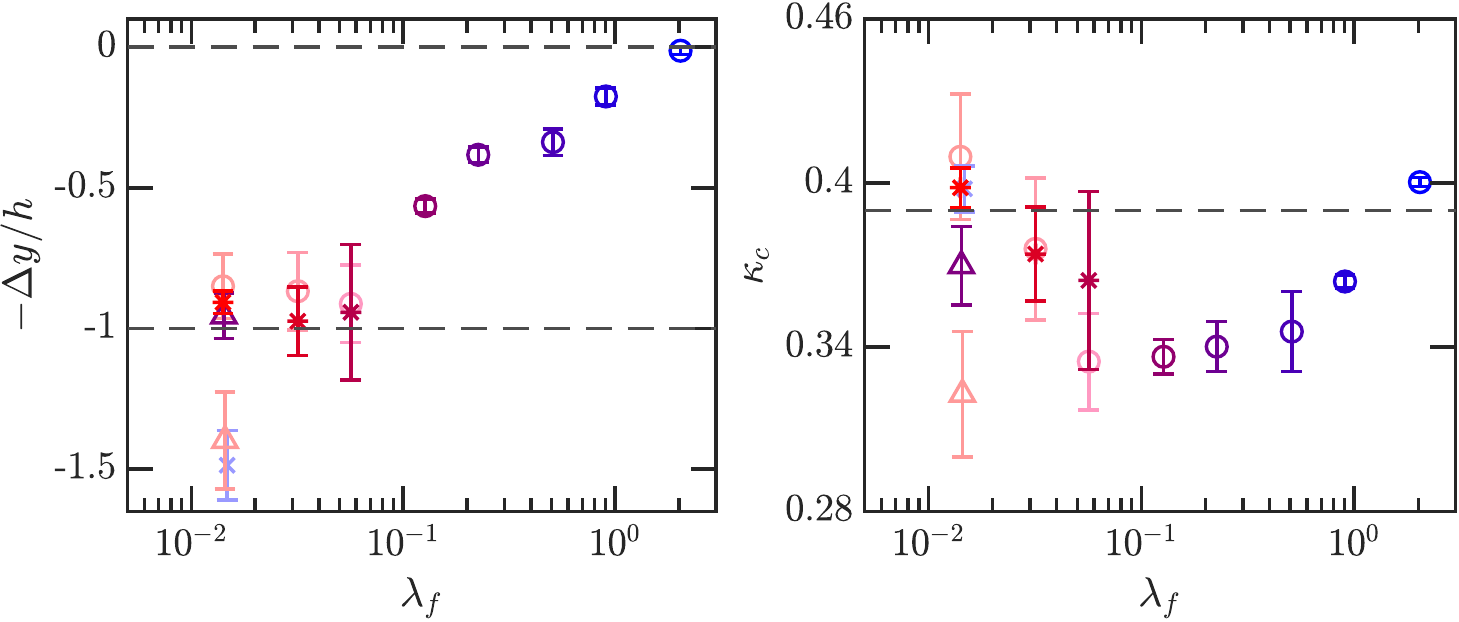}
    \put (-137mm,56mm) {$(a)$}
    \put (-67mm,56mm) {$(b)$}
    \caption{ $(a)$ Ratio of zero-plane-displacement depth to canopy height, $-\Delta y/h$, and $(b)$ Kármán constant, $\kappa_c$, versus canopy density, $\lambda_f$, for all cases considered. Markers $\bigcirc$, $\ast$, $\times$, and $\Delta$ represent full-channel cases at $Re_\tau\approx550$ and $Re_\tau\approx900$ and open-channel cases at $Re_\tau\approx550$ and $Re_\tau\approx1000$, respectively. The colour scheme is as in figure \ref{fig:cond_stats}. Transparent markers represent cases with substantial texture-induced coherent flows in the outer layer, and coloured markers represent cases with a smooth-wall-like outer layer. In $(a)$, \protect\dashedk represents $\Delta y$ corresponding to zero-plane-displacement height at the canopy tip ($-\Delta y/h=0$) and at the floor ($-\Delta y/h=-1$). In $(b)$, \protect\dashedk represents $\kappa_s\approx0.39$.}
    \label{fig:dy_kappa}
\end{figure}

For the cases with less intrusive canopies, C36$_{550}$ to C144$_{550}$, C216$_{900}$ to C432$_{900}$, and O400$_{1000}$, the r.m.s. velocity fluctuations, Reynolds shear stress and pre-multiplied spectral energy densities scaled with $u_\tau^*$ essentially collapse with their respective smooth-wall values above a height $y_*\approx0.3\delta$, as shown in figure \ref{fig:rms}, \ref{fig:uvm} and \ref{fig:sp_03}. This suggests that $u_\tau^*$, and not $u_\tau^\star$, is the velocity scale for the outer-layer turbulence. However, for the more intrusive cases, for example O800$_{1000}$, the roughness sublayer extends well into the overlying flow, which no longer exhibits a smooth-wall-like logarithmic layer, as evidenced by the footprint of the element-induced flow in figure \ref{fig:sp_03}$(i-l)$. These intrusive textures cause a more in-depth modification of the outer-layer flow, and can lead to the breakdown of outer-layer similarity. Similar observations have been reported on large riblets and porous walls, where outer-layer similarity is limited if surface effects penetrate into a significant portion of $\delta$ \citep{breugem2006influence, manes2011turbulent, endrikat2022reorganisation}.

By minimising the r.m.s. deviation of $\beta_c$ from $\beta_s$, as discussed in \S \ref{subsec:Sens} \& \ref{subsec:Univ}, we obtain the values of $\Delta y$ and $u_\tau^*$ that provide the scales for the outer-layer turbulent flow. The trend of the zero-plane displacement in figure \ref{fig:dy_kappa}$(a)$ suggests that the flows over dense canopies perceive an origin close to the tips, and that this origin becomes deeper as the canopy density decreases. In figure \ref{fig:dy_kappa}$(b)$, the densest case has a smooth-wall-like Kármán constant, $\kappa_c\approx0.4$, and the value of $\kappa_c$ decreases as the canopy density decreases. However, for the sparsest cases, $\kappa_c$ appears to tend back to its smooth-wall value. We note that the decreases in $\kappa_c$ for all canopies in this study never exceeds $15\%$ of its smooth-wall value, which is significantly smaller than obtained by fitting $U^+$ to a $\log(y_*^+)$ profile as in \cite{breugem2006influence}, \cite{suga2010effects}, \cite{rosti2017numerical}, \cite{kuwata2017direct}, \cite{kazemifar2021effect}, and \cite{okazaki2021describing, okazaki2022turbulent}, where the values of $\kappa_c$ reported were as low as \colcomm{$\kappa_c\approx0.2$.}

\colcomm{The} present results for dense and sparse canopies are consistent with the results in \cite{nepf2012flow}, \cite{brunet2020turbulent} and \cite{chung2021predicting}, who summarised that for roughness in the dense or sparse regime, the zero-plane-displacement height approaches the roughness crest or trough, respectively. For instance, for the dense canopy C36$_{550}$, $\lambda_f\approx2.04$, the smooth-wall-like overlying flow perceives an origin at the tips, and the Kármán constant has a smooth-wall value, $\kappa_c\approx\kappa_s\approx0.39$, as illustrated in figure \ref{fig:dy_kappa}. \cite{macdonald2016turbulent} and \cite{sharma2020turbulent_b} have also observed such skimming flow for closely-pack sinusoidal roughness and filament canopies, where $s^+\lesssim3$. Particularly, in the dense regime, where the spacing between elements is comparable to the viscous length scale, $\nu/u_\tau$, the overlying turbulence is essentially precluded from penetrating the roughness, and the unmodified overlying flow is smooth-wall-like with a zero-plane-displacement height at the tips.

Sparse canopies, where the ratio between element spacing and height is large, are generally more intrusive than the intermediate to dense canopies because the roughness sublayer can extend to $y/s\approx1$ into the channel, as discussed in \S \ref{subsec:Extent}. For the sparse and tall canopies, C216$_{550}$, C288$_{550}$, C432$_{550}$, O400$_{550}$ and O800$_{1000}$, the surface effects penetrate deep into the overlying flow and can reach the channel centre, impeding the assessment of outer-layer similarity. For these substrates, outer-layer similarity could only be assessed at higher $Re_\tau$. This motivated us to conduct simulations C216$_{900}$, C288$_{900}$, C432$_{900}$, O400$_{1000}$, for which the canopies have similar dimensions to C216$_{550}$, C288$_{550}$, C432$_{550}$, O400$_{550}$, in inner units, but such that the unperturbed core flow is a larger portion of the channel. As illustrated in figure \ref{fig:dy_kappa}, the larger-$Re_\tau$ flows allow for a clearer assessment of outer-layer similarity, with zero-plane-displacement height at the floor and $\kappa_c\approx\kappa_s\approx0.39$. Our observations are consistent with those in \cite{poggi2004effect}, \cite{nepf2007retention} and \cite{sharma2020scaling}, where the flows over sparse canopies exhibit characteristics of smooth-wall flows, with reference wall at the canopy floor.

Flows over canopies with an intermediate density ($\lambda_f\gtrsim0.1$) perceive an origin between the tips and floor, as shown in figure \ref{fig:dy_kappa}$(a)$. For these intermediate canopies, with $s/h\lesssim1$, the overlying flow interacts mainly with the upper part of the obstacles and turbulence does not penetrate all the way to the floor \citep{grimmond1999aerodynamic, luhar2008interaction, macdonald2018direct, sharma2020turbulent_b}. These intermediate-canopy flows exhibit values of the Kármán constant $\kappa_c\approx0.34-0.36$, different from the smooth-wall value $\kappa_s\approx0.39$, implying that the intermediate canopies disrupt the overlying flow more profoundly than both dense and sparse substrates. Nevertheless, other than for the change in $\kappa$, the turbulent statistics remain essentially smooth-wall-like in the logarithmic layer and above, as depicted in figures \ref{fig:df}, \ref{fig:diag}, and \ref{fig:rms}-\ref{fig:sp_03}. This suggests a modified outer-layer similarity, where $\kappa_c\neq0.39$, but where otherwise the turbulence is outer-layer-similar to a smooth-wall flow. For $h\ll\delta$, we could expect that the flow far from the wall is not influenced by the details of the surface topology, as is the classical view of wall turbulence \citep{clauser1956turbulent}. Further work is required to study if the same canopies would also exhibit $\kappa_c\neq0.39$ at larger $\delta/h$ ratios, say $\delta/h\approx40$ as proposed by \cite{jimenez2004turbulent}.

\section{Conclusions}\label{sec:Conclusions}
In the present work, we have assessed outer-layer similarity in flows over canopies ranging from sparse to dense ($\lambda_f\approx0.01-2.03$) at $Re_\tau\approx550-1000$. We have discussed and appraised the conventional methods for the assessment of outer-layer similarity, showing that these methods could be inaccurate for flows over dense roughness and for flows at moderate $Re_\tau$.

To investigate outer-layer turbulence, we first determined the depth of the roughness sublayer, within which the turbulence is significantly disrupted by the element-induced flow. It was shown that the roughness sublayer for the present flows extends to a height equal to the element spacing. For the cases with tall ($h\approx0.2\delta$) and sparse ($\lambda_f\ll0.1$) elements, C216$_{550}$, C288$_{550}$, C432$_{550}$, O400$_{550}$, and O800$_{1000}$, the element-induced flows penetrate effectively into the channel as far as $y\approx0.5\delta$, or even above. As a result, these intrusive canopies leave only a small portion of core flow unperturbed, making the assessment of outer-layer similarity difficult. To verify whether outer-layer similarity can recover for these intrusive cases, simulations at higher $Re_\tau$, producing a larger unperturbed region, where required.

Conventionally, outer-layer similarity for a flow over roughness is recovered by imposing \colcomm{a zero-plane displacement and the corresponding friction velocity} on the mean-velocity profile, such that the logarithmic layer is smooth-wall-like. \colcomm{We have discussed some caveats of} conventional drag-based and mean-velocity-based methods used to determine \colcomm{these constants. For drag-based methods, and for densely packed roughness elements, the point of action of the total drag is located within the elements,} while the overlying flow actually perceives an origin in the vicinity of the tips.  For mean-velocity-based methods, we have shown that \colcomm{at moderate $Re_\tau$ even smooth-wall flows do not exhibit a constant slope for $U^+$ v.s. $\log(y_*^+)$, or a plateau in $\beta$. Therefore,} matching the shape of $U^+$ to a $\log(y_*^+)$ or, equivalently, enforcing a plateau in the diagnostic function, $\beta$, \colcomm{may result in an artificial breakdown of outer-layer similarity and spurious predictions for the zero-plane displacement and the friction velocity at any but the highest $Re_\tau$ available in DNS literature.}

To obtain the zero-plane displacement and friction-velocity scale that recovers a smooth-wall-like diagnostic function, we minimise the deviation between the canopy and smooth-wall diagnostic function everywhere above the roughness sublayer, instead of in the logarithmic layer alone. By this method, we obtain a smooth-wall-like $\beta$ not only in the logarithmic layer, but also in the wake region, enforcing outer-layer similarity in its full sense. We also explore the possibility of the zero-plane displacement and the friction velocity being set independently, but find that outer-layer similarity is more consistently recovered when they are coupled. For the dense canopies like C36$_{550}$, the unmodified overlying flow is smooth-wall-like with a zero-plane-displacement height at the tips, and the Kármán constant has a smooth-wall value $\kappa_c\approx\kappa_s\approx0.39$. This is the case because the skimming turbulence is precluded from penetrating into the obstacles, and perceives an origin at the tips. For sparse canopies ($\lambda_f\ll0.1$), the higher-$Re_\tau$ cases,  C216$_{900}$, C288$_{900}$, C432$_{900}$, and O400$_{1000}$, allow for the assessment of outer-layer similarity, with zero-plane-displacement height at the floor, and $\kappa_c\approx\kappa_s\approx0.39$. For intermediate canopies ($\lambda_f\gtrsim0.1$), with $s/h\lesssim1$, the overlying flow interacts mainly with the upper part of the obstacles, and turbulence does not penetrate all the way to the floor. These intermediate-canopy flows perceive a zero-plane-displacement height between the tips and floor, and exhibit values of the Kármán constant $\kappa_c\approx0.34-0.36$ different from the smooth-wall value $\kappa_s\approx0.39$, implying that the intermediate canopies disrupt the overlying flow more profoundly than both dense and sparse substrates. Nevertheless, other than for the change in $\kappa$, the turbulent statistics remain essentially smooth-wall like in the logarithmic layer and above.

\section*{Acknowledgements}
This work was partially supported by EPSRC grant EP/S013083/1. For the purpose of open access, the authors have applied a Creative Commons Attribution (CC BY) licence to any Author Accepted Manuscript version arising from this submission. Computational resources were provided by the ‘Cambridge Service for Data Driven Discovery’ operated by the University of Cambridge Research Computing Service and funded by EPSRC Tier-2 grant EP/P020259/1 (projects: cs066 and cs155), and by the 'ARCHER2' system in the UK, funded by PRACE (project: pr1u1702) and EPSRC (project: e776). All research data supporting this publication is directly available within this publication.

\section*{Declaration of interests}
The authors report no conflict of interest.

\appendix
\section*{Appendix. Grid independence and validation}\label{appA}

\begin{figure}
    \vspace*{3mm}
    \centering
    \includegraphics[width=0.65\textwidth]{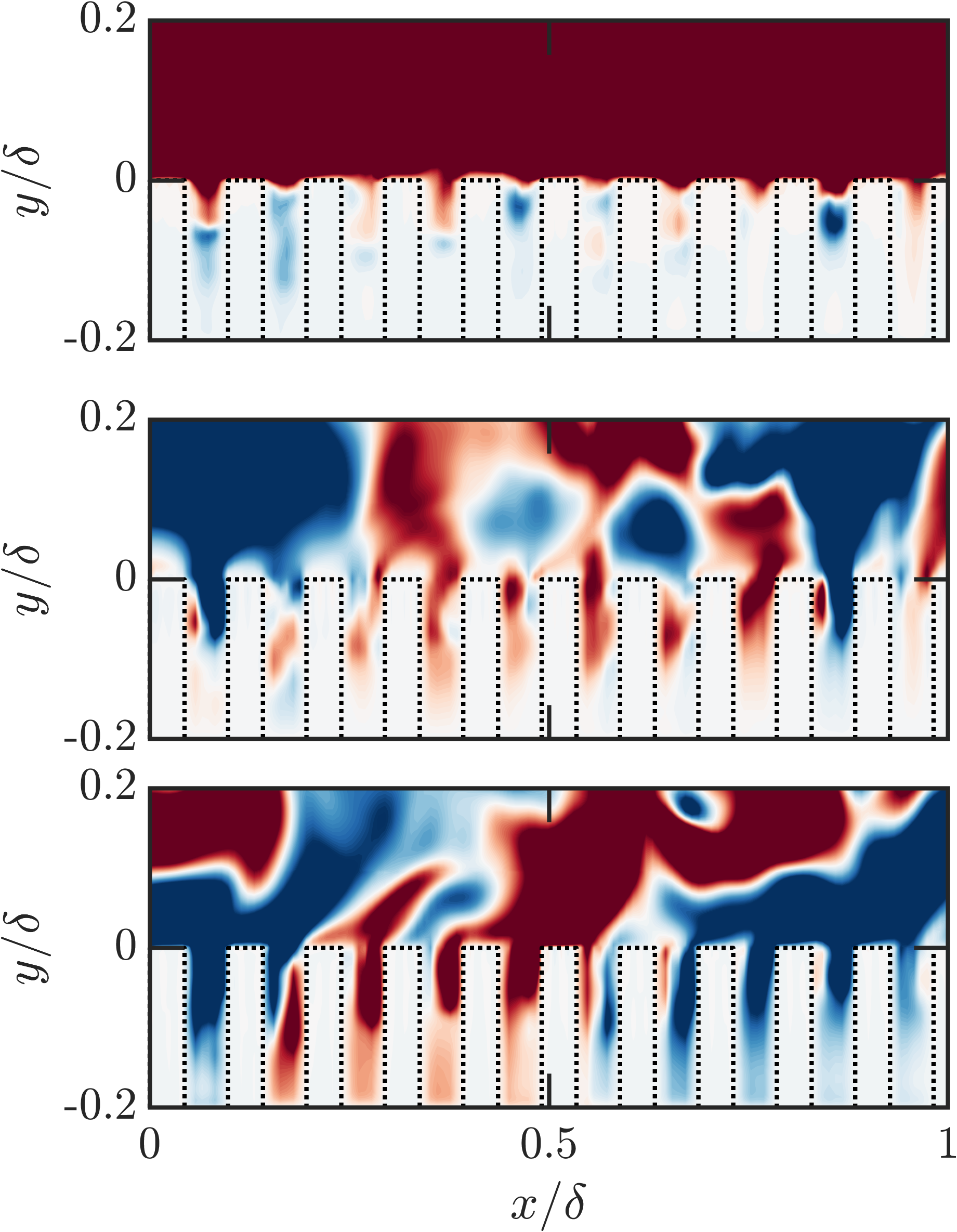}
    \put (-88mm,110mm) {$(a)$}
    \put (-88mm,73.5mm) {$(b)$}
    \put (-88mm,39.5mm) {$(c)$}
    \caption{Instantaneous realisations of the (a) streamwise, (b) wall-normal and (c) spanwise velocities in a plane intersecting the canopy elements for case C54$_{550}$, scaled with $u_\tau$ evaluated at the canopy tips. The darkest red and blue contours represent intensities of $\pm1.0$, respectively.}
    \label{fig:vel_validate}
\end{figure}

In this study, the no-slip condition within the canopy elements is enforced using the direct-forcing, immersed-boundary method \colcomm{\citep{iaccarino2003immersed} as} implemented in \cite{garcia2011hydrodynamic} and modified by \cite{sharma2020scaling, sharma2020turbulent_b}. This method applies a body force term to the right-hand side of equation (\ref{eq:ns_dis1}) and drives the velocity at the immersed-boundary points to zero. The reader is referred to the works of \cite{sharma2020turbulent_b} and \cite{sharma2020turbulent} for a detailed discussion on the accuracy of the immersed-boundary method implemented and wall-parallel resolution used. In the present study, the velocity within the rigid canopy elements is observed to be less than $0.1u_\tau$\colcomm{, with $u_\tau$ evaluated at the canopy tips,} for all the DNSs conducted. As portrayed in figure \ref{fig:vel_validate}, the velocity at the 'solid' points within the obstacles is significantly smaller than that at the surrounding 'fluid' points, illustrating that the immersed-boundary method resolves the canopy topology.

\begin{figure}
    \vspace*{5mm}
    \centering
    \includegraphics[width=0.5\textwidth]{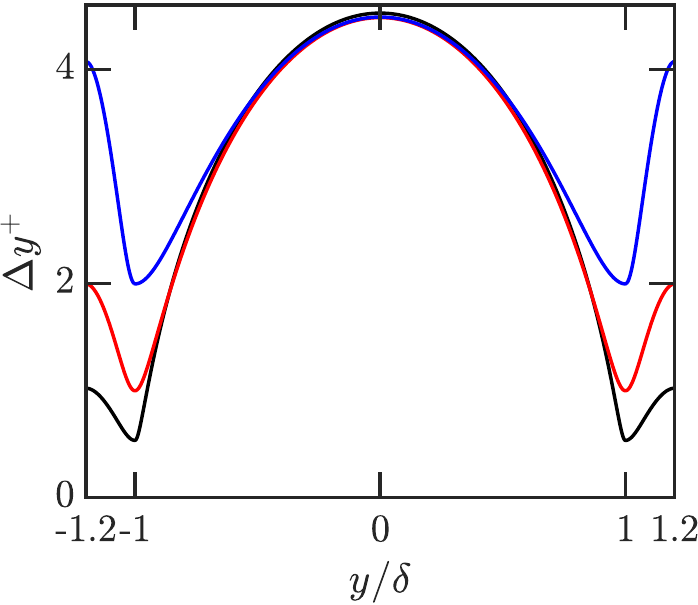}
    \vspace*{2mm}
    \caption{Wall-normal resolution, scaled with $\nu$ and $u_\tau$ evaluated at the canopy tips, for case C108$_{550}$ for a fine resolution \protect\solidk, a standard resolution \protect\solidr, and a coarse resolution \protect\solidb.}
    \label{fig:grid_size}
\end{figure}

\begin{figure}
    \vspace*{7mm}
    \centering
    \includegraphics[width=\textwidth]{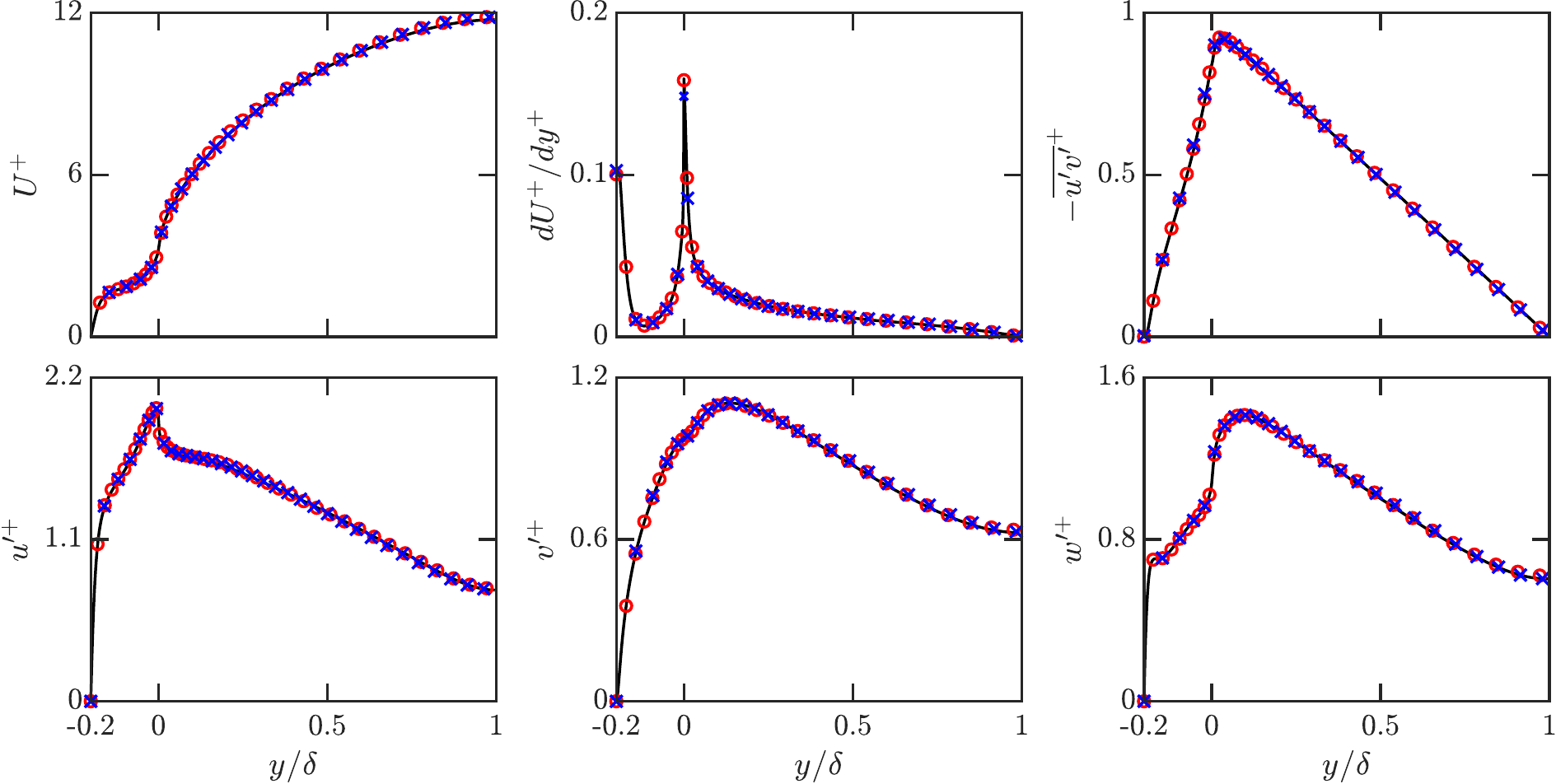}
    \put (-136mm,66mm) {$(a)$}
    \put (-91mm,66mm) {$(b)$}
    \put (-45mm,66mm) {$(c)$}
    \put (-136mm,34mm) {$(d)$}
    \put (-91mm,34mm) {$(e)$}
    \put (-45mm,34mm) {$(f)$}
    \vspace*{2mm}
    \caption{Turbulent statistics based on $\nu$ and $u_\tau$ evaluated at the canopy tips for case C108$_{550}$. The colour scheme is as in figure \ref{fig:grid_size}.}
    \label{fig:grid_rms}
    \vspace*{2mm}
\end{figure}
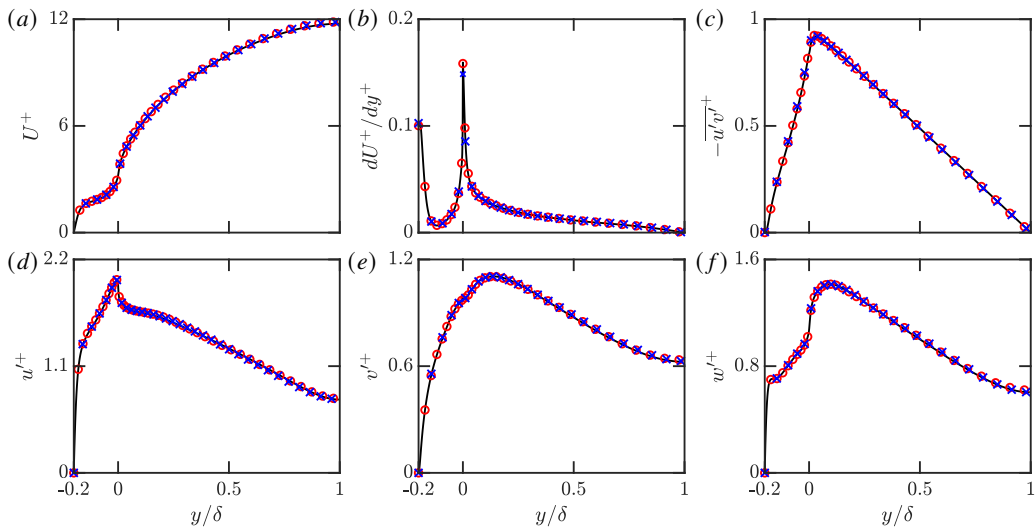

To analyse grid convergence, we have carried out three DNSs for case C108$_{550}$ with different wall-normal resolutions, as portrayed in figure \ref{fig:grid_size}. Compared to smooth-wall flows, where the maximum mean shear occurs at the wall, $dU^+/dy^+=1$, in our flows it occurs at the canopy-tip plane and is not as high, in the present case $dU^+/dy^+\approx0.16$, peaking again at the floor at $dU^+/dy^+=0.1$. Since the local shear is the driver of turbulence, the wall-normal resolution required for the direct simulation of these flows is thus lower than for smooth-wall flows in terms of $\Delta y^+$, and can be adjusted following the local shear, as shown in figures \ref{fig:grid_size} and \ref{fig:grid_rms}$(b)$. Consistent with this discussion, let us also note that the resolution at the floor in all of our simulations, for the case portrayed (C108$_{550}$) $\Delta y^+\approx2$, is actually $\Delta y^+\approx0.3$ when scaled with the local, dynamically relevant friction velocity proposed in \cite{sharma2018turbulent, sharma2020turbulent_b}. Figure \ref{fig:grid_rms} portrays the results obtained for the resolution used in this paper plus one finer and one coarser. All results collapse, evidencing grid convergence.


\bibliographystyle{jfm}
\bibliography{outlaysim}

\end{document}